\documentclass[12pt]{article}
\usepackage[utf8]{inputenc}
\usepackage[T1]{fontenc}
\usepackage[a4paper,margin=2.5cm]{geometry}
\usepackage{amsmath,amssymb,amsthm,mathtools}
\usepackage{physics}
\usepackage{graphicx}
\usepackage{subcaption}
\usepackage{bm}
\usepackage{cite}
\usepackage{hyperref}
\hypersetup{colorlinks=true,linkcolor=blue,citecolor=blue,urlcolor=blue}
\usepackage{microtype}
\usepackage{authblk}
\usepackage{tikz}
\usetikzlibrary{arrows.meta}

\numberwithin{equation}{section}

\title{On the Dynamics of Multiparticle Carroll--Schr\"odinger Quantum Systems}

\author[1]{Jos\'e Rojas}
\author[1,2]{Melvin Arias}

\affil[1]{\textit{\small Instituto de F\'isica, Universidad Aut\'onoma de Santo Domingo, Av. Alma Mater, Santo Domingo 10105, Dominican Republic}}
\affil[2]{\textit{\small Laboratorio de Nanotecnolog\'ia, \'Area de Ciencias B\'asicas y Ambientales, Instituto Tecnol\'ogico de Santo Domingo, Av. Los Pr\'oceres, Santo Domingo 10602, Dominican Republic}}

\date{\today}

\begin{document}

\maketitle
\begin{abstract}
We study the dynamics of multiparticle Carroll--Schr\"odinger (CS) quantum systems in $1{+}1$ dimensions, where $x$ acts as the evolution variable and $t$ as the configuration coordinate. We derive the $N$-body theory on equal-$x$ slices as the Carrollian limit of a relativistic multi-time Klein--Gordon model, introducing temporal interactions via minimal coupling to the temporal energy operators. An $x$-dependent gauge transformation maps this to an equivalent description with explicit many-body potentials, illustrated by a temporal coupled-oscillator model that exhibits synchronization. Adopting a complementary spatial viewpoint with a static potential $U_{\!tot}(\mathbf x)$, we show that the evolution is driven by the collective force $\sum_j\partial_{x_j}U_{\!tot}$; for any translation-invariant interaction (such as a regularized Coulomb potential), these internal forces cancel, rendering the collective dynamics free and highlighting Carrollian ultralocality. We also construct a coordinate duality mapping separable Schr\"odinger Hamiltonians to CS generators via Schwarzian derivatives. Exchange symmetry is formulated in the time domain, yielding temporal bunching for bosons and antibunching for fermions via the second-order coherence function $g^{(2)}(t,t')$. In second quantization, the contact limit yields a temporal derivative cubic--quintic nonlinear Schr\"odinger equation with a theoretically fixed nonlinearity coefficient $\beta=-3/16$. Finally, by coupling canonical pairs to external scalar and gauge fields, we establish an isomorphism with one-dimensional current-density functional theory, outlining a Carrollian Hohenberg--Kohn mapping and Kohn--Sham scheme.
\end{abstract}

\clearpage

\section{Introduction}

Nonrelativistic quantum mechanics on the line is usually formulated in the Schr\"odinger picture, where time $t$ plays the role of evolution parameter and $x$ is the configuration coordinate. The opposite corner of kinematical limits is the \emph{Carroll} regime, obtained by a $c\!\to\!0$ contraction of the Poincar\'e group \cite{LevyLeblond1965}. At the level of wave equations, this limit can be implemented by the Carroll--Schr\"odinger (CS) equation \cite{Najafizadeh2025-1,Najafizadeh2025-2}
\begin{equation}
 i\hbar c\,\partial_x\Psi(x,t)\;-\;
 \frac{1}{2mc^2}\big(-i\hbar\partial_t\big)^2\Psi(x,t)=0,
 \label{eq:CS1}
\end{equation}
which is first order in the spatial variable $x$ (the evolution coordinate) and second order in the temporal configuration coordinate $t$. This admits a unitary $x$-evolution on the equal-$x$ Hilbert space $\mathcal H=L^2(\mathbb R_t,dt)$ generated by
\begin{equation}
 H_x=\frac{\widehat E^{\,2}}{2mc^2},\qquad \widehat E:=-i\hbar\,\partial_t,
 \label{eq:Hx}
\end{equation}
self-adjoint on $H^2(\mathbb R_t)$ \cite{Rojas2025}. In earlier work we used this operator picture to relate Schr\"odinger and CS dynamics at the single-particle level, including currents, dispersion and classical limits, and to construct coordinate transformations connecting static Schr\"odinger potentials to time-dependent Carroll potentials, and to introduce the equal-$x$ Hilbert space together with a proof of unitary evolution in $x$ for the one-particle CS wave function \cite{Rojas2025}.

The aim of this paper is to extend the framework to \emph{many-body} systems and to examine, in a controlled setting, how interactions, correlations and coordinate dualities act when the roles of $x$ and $t$ are exchanged. We work on the equal-$x$ Hilbert space $\mathcal H_N=L^2(\mathbb R^N_{\bm t})$ and introduce temporal interactions by minimally coupling one-body fields and two-time kernels to the temporal energy operators $\widehat E_i$. An $x$-dependent many-body gauge transformation produces an equivalent evolution equation in which the interaction appears as an explicit $N$-body potential in the time variables. For gauge potentials that are affine in $x$, this can be organized so that the resulting temporal potential has a prescribed form. This provides a simple mapping: starting from a standard many-body Schr\"odinger model one formally replaces the spatial coordinates $x_i$ by times $t_i$ to obtain a candidate Carrollian partner whose equal-$x$ dynamics has the same $T{+}V$ structure but acts along $t$. As an illustration we discuss a temporal analogue of coupled harmonic oscillators, where the temporal configuration variables have the same normal-mode structure as in the spatial model.

We also adopt a complementary, purely spatial viewpoint and start from a static total potential energy $U_{\!tot}(\mathbf x)$. Writing the CS evolution with $U_{\!tot}$ as a shift of the temporal energy operator and performing a gauge transformation in $t$, we show that, in this construction, the Carrollian dynamics depends on $U_{\!tot}$ through the effective force on the center of mass $\sum_j\partial_{x_j}U_{\!tot}(\mathbf x)$. For coupled harmonic oscillators on a line this produces a linear-in-$t$ drive of the collective coordinate, and, in order to illustrate the resulting dynamics, we obtain explicit Gaussian $N$-body solutions while showing that the multiparticle wave function's spatial distribution and evolution are highly dependent on the boundary data, discussing their physical interpretation from the Carrollian point of view. For a regularized one-dimensional Coulomb interaction (and translation-invariant potentials generally), by contrast, the internal forces cancel in the collective direction and the transformed dynamics is free, with the interaction potential reappearing only through phases and boundary data, highlighting an important limitation of multiparticle Carroll--Schr\"odinger systems with purely spatial and translation-invariant internal interactions.

A second theme of the paper is a many-body generalization of the coordinate duality between the time-independent Schr\"odinger equation and a space-independent CS equation that was previously analyzed in the one-body setting \cite{Rojas2025}. Restricting, on the Schr\"odinger side, to separable $N$-body Hamiltonians with $\sum_i V_{\mathrm{sch}}^{(i)}(x_i)$, we construct, for each particle, a potential-dependent reparametrization $x_i=\delta_i(t_i)$ that maps the stationary Schr\"odinger problem to a space-independent CS equation in the time variables with one-body Carroll potentials $V_{\mathrm{car}}^{(i)}(t_i)$. The map is governed by Schwarzian relations for the inverse functions $\tau_i=\delta_i^{-1}$, and provides an explicit relation between families of static Schr\"odinger potentials and space-independent Carroll potentials acting on the temporal coordinates. This construction extends the single-particle duality of Ref.~\cite{Rojas2025} to a separable many-body setting and clarifies the role of the Schr\"odinger energies $E_{\mathrm{sch}}^{(i)}$ and Carroll momenta $p_0^{(i)}$ as dual labels, and we explore the Schwarzian coordinate map in an interacting example of two Schr\"odinger coupled harmonic oscillators.

Once the many-body generators are in place, exchange symmetry can be formulated directly in the time domain. On each equal-$x$ slice the $N$-body wave function lives in $L^2(\mathbb R^N_{\bm t})$, and indistinguishability acts on permutations of the time labels. This leads to bosonic and fermionic sectors \emph{in time}, defined by symmetry or antisymmetry under $t_i\leftrightarrow t_j$ at fixed $x$. The associated one-body density matrix and pair density quantify temporal coherence and Pauli suppression on the coincidence hyperplanes $t_i=t_j$. In this setting it is natural to introduce a second-order coherence function $g^{(2)}(t,t')$ on equal-$x$ slices and to interpret it as a temporal analogue of Hanbury Brown--Twiss (HBT) correlations. In the noninteracting reference cases we recover temporal bunching for bosons and antibunching for fermions.

The many-body framework also supports effective field and functional descriptions. In second quantization on equal-$x$ slices we use canonical commutation relations in the time coordinate \cite{Najafizadeh2025-2} to derive a temporal nonlinear Schr\"odinger equation with a two-time interaction kernel. In a short-memory limit, analogous to the short-range interaction limit of a one-dimensional Lieb--Liniger gas~\cite{LiebLiniger1963I,LiebLiniger1963II}, where the kernel becomes local in $t$, the corresponding mean-field equation reduces, after simple rescalings, to a derivative cubic--quintic nonlinear Schr\"odinger equation (DNLS) in the time coordinate. We find that the quintic nonlinearity coefficient is geometrically fixed to $\beta=-3/16$, placing the interaction-driven temporal Carroll gas in a specific universality class of DNLS systems.

Finally, we transpose density-functional concepts to the CS setting in a static (equal-$x$) framework. By minimally coupling the canonical momentum $P$ to a scalar field $\Phi(t)$ and the temporal energy $E$ to a gauge field $U(t)$, we identify a natural mapping that renders the ground-state problem formally isomorphic to one-dimensional current-density functional theory (CDFT). Relying on this isomorphism, we introduce a universal functional $F[n,j_t]$ of the temporal density $n(t)$ and temporal current $j_t(t)$ and outline a Carrollian Kohn--Sham scheme in terms of one-time orbitals that reproduce $(n,j_t)$ on equal-$x$ slices.

The remainder of the paper is organized as follows. In Sec.~\ref{section2} we obtain the many-body CS formalism as the limit of a relativistic multi-time Klein--Gordon system and describe a mapping to explicit temporal potentials, illustrated by a temporal coupled-oscillator model. In Sec.~\ref{sec:space-viewpoint} we explore the spatial interactions viewpoint and analyze model systems of coupled harmonic oscillators and systems with Coulomb interactions, finding that translation-invariant internal spatial forces cancel in the collective Carrollian dynamics. The following section extends the single-particle coordinate duality between the time-independent Schr\"odinger equation and the space-independent CS equation to separable $N$-body systems. Section~\ref{sec:time-exchange-HBT} discusses exchange symmetry in the time domain and re-interprets temporal HBT correlations from a Carrollian perspective. The next section obtains a temporal nonlinear Schr\"odinger equation in second quantization; in the contact limit, this reduces to a derivative cubic--quintic NLSE with a geometrically determined coefficient. Finally, Section~\ref{sec:Carroll-HK-KS} outlines a proposal for a Carrollian density-functional framework. By minimally coupling the canonical momentum $P$ to a scalar field $\Phi$ and the energy $E$ to a gauge field $U$, we identify a formal isomorphism with one-dimensional current-density functional theory, which motivates a Carrollian Hohenberg--Kohn mapping and Kohn--Sham scheme. We conclude with a brief discussion of implications and limitations.

\section{Many-body Carroll--Schr\"odinger temporal formalism}
\label{section2}

To motivate the many-body Carroll--Schr\"odinger (CS) generator, it is
useful to start from a relativistic parent theory. We work in $1{+}1$
dimensions and consider an $N$-particle Klein--Gordon
field
\[
\Phi(\bm q,\bm t)\equiv\Phi(q_1,\ldots,q_N;\,t_1,\ldots,t_N),
\]
where each particle is described by its own space--time pair
$(q_i,t_i)$, following Dirac’s relativistic multi-time formulation
\cite{Dirac1932}. For each particle $i$ we impose a Klein--Gordon
equation in its own time coordinate $t_i$,
\begin{equation}
\bigg(
  -\,\frac{1}{c^{2}}\partial_{t_i}^{2}
  + \partial_{q_i}^{2}
  + \mu^{2}
\bigg)\Phi(\bm q,\bm t)=0,
\qquad
\mu:=\frac{mc}{\hbar},
\qquad
i=1,\ldots,N,
\label{eq:multi-KG-compact}
\end{equation}
which is the natural many-body generalization of the one-body
tachyonic Klein--Gordon equation used in Ref.~\cite{Najafizadeh2025-2}.
Equations of the form \eqref{eq:multi-KG-compact} also fit into the
multi-time-wave-function framework reviewed in
Ref.~\cite{LienertPetratTumulka2017}, but here we use them only as a
relativistic parent for the Carroll contraction.

In a general relativistic setting, the condition of ``co-spatiality''
($q_1=\cdots=q_N=x$) is frame-dependent, much like simultaneity
($t_1=\cdots=t_N=t$). However, we are specifically interested in the
Carrollian limit ($c \to 0$), where the spacetime lightcone collapses
and spatial intervals become absolute while temporal intervals become
relative. In this limit, the equal-$x$ foliation becomes
Carroll-invariant (since $x' \approx x$ under Carroll boosts).
Therefore, we perform the reduction to equal-$x$ slices in the
reference frame where the Carroll contraction is defined, identifying
it as the natural configuration space for the limiting theory.

Restricting to these equal-$x$ slices and following the same field
redefinition and Carroll scaling procedure as in
Ref.~\cite{Najafizadeh2025-2}, one finds that each pair $(x,t_i)$
produces a one-body CS operator $\widehat E_i^{\,2}/(2mc^{2})$ acting
on the equal-$x$ wave function $\Psi(x,\bm t)$. Consistency of the
$x$-evolution then identifies the total generator as the sum of these
one-body pieces, leading to the free many-body CS equation
\begin{equation}
  i\hbar c\,\partial_x\Psi(x,\bm t)
  =
  H_x^{(N)}\,\Psi(x,\bm t)
  =
  \sum_{i=1}^{N}\frac{\widehat E_i^{\,2}}{2mc^2}\,\Psi(x,\bm t),
  \label{eq:CS-N-free}
\end{equation}
with
\begin{equation}
  H_x^{(N)}
  :=
  \sum_{i=1}^N \frac{\widehat E_i^{\,2}}{2mc^2},
  \qquad
  \widehat E_i := -i\hbar\,\partial_{t_i}.
  \label{eq:HN-free}
\end{equation}
Here $m$ is the single-particle mass; $H_x^{(N)}$ is simply the direct
sum of $N$ identical one-body Carroll generators, in exact analogy with
the usual Schr\"odinger Hamiltonian
$H_{\mathrm{Sch}}^{(N)}=\sum_i p_i^{2}/(2m)$.

Although \eqref{eq:multi-KG-compact} is a multi-time system in the
sense of Dirac and of Ref.~\cite{LienertPetratTumulka2017}, the
resulting Carroll theory is \emph{not} a multi-time evolution in the
Schr\"odinger sense. After the Carroll contraction we obtain a single
first-order equation in $x$, with $t_1,\ldots,t_N$ appearing only as
configuration coordinates in the Hermitian generator $H_x^{(N)}$. There
are no separate evolution equations $i\hbar\partial_{t_i}\Psi=H_i\Psi$
and therefore no compatibility conditions of the type that lead to the
multi-time no-go theorems \cite{PetratTumulka2014}. In this work we
interpret \eqref{eq:CS-N-free} as defining a standard unitary evolution
in $x$ on equal-$x$ slices.

\medskip

On each equal-$x$ slice the state therefore lives in the Hilbert space
\begin{equation}
  \mathcal H_N := L^2(\mathbb R^N_{\bm t},d\bm t),
  \qquad
  \bm t = (t_1,\ldots,t_N),
\end{equation}
and \eqref{eq:CS-N-free} describes unitary $x$-evolution on $\mathcal
H_N$. The temporal variables $(t_1,\ldots,t_N)$ play the role of
configuration coordinates, so that $|\Psi(x,\bm t)|^{2}$ is a genuine
probability density in the time domain on each equal-$x$ slice. Among
the various Carrollian viewpoints, this equal-$x$ picture is singled
out by providing a linear, norm-preserving dynamics with a
positive-definite density; in Sec.~\ref{sec:space-viewpoint} we will
contrast it with a complementary spatial perspective.

\subsection{Temporal interactions}

The free generator \eqref{eq:HN-free} can be generalized to include
interactions in several ways. A relativistically natural option,
following the dynamic-time models of
Refs.~\cite{Horwitz1996,Horwitz2015}, is
to introduce potentials that depend on Lorentz-invariant separations
\[
  \rho_{ij}^{2}
  := c^{2}(t_i-t_j)^{2} - (q_i-q_j)^{2},
\]
so that the Klein--Gordon equation acquires a nonlocal but
Lorentz-invariant interaction $V(\{\rho_{ij}\})$. On equal-$x$ slices
$q_i=q_j=x$ the invariant separations reduce to purely temporal
distances,
\[
  \rho_{ij}^{2}\big|_{q_i=q_j=x}
  = c^{2}(t_i-t_j)^{2},
\]
and the interaction depends only on differences of the time
coordinates,
\[
  V(\{\rho_{ij}\})\Big|_{q_i=q_j=x}
  = V\big(\{c|t_i-t_j|\}\big)
  \equiv V_t(\bm t).
\]
Taking the Carroll limit as above then produces temporal interaction
terms $V_t(\bm t)$ acting on $\Psi(x,\bm t)$, in addition to the free
generator $H_x^{(N)}$. This provides a relativistic origin for the
purely temporal many-body potentials that we will use in this section.

A complementary construction, analyzed in
Sec.~\ref{sec:space-viewpoint}, starts instead from a static spatial
potential energy $U_{\!tot}(\bm x)$ and leads, after a gauge
transformation and Carroll limit, to an effective drive governed by the
collective spatial force $\sum_j\partial_{x_j}U_{\!tot}(\bm x)$. This
spatial viewpoint is useful for comparing with standard Schr\"odinger
models, but yields rather constrained collective dynamics in strictly
translation-invariant cases.

For the temporal viewpoint adopted here, it is convenient to parametrize
interactions directly on $\mathcal H_N$ in an
\emph{inside-the-square} form by
replacing each temporal energy operator $\widehat E_i$ with a minimally
coupled operator $\widehat E_i - A_i(x;\bm t)$:
\begin{equation}
  \mathcal H^{(N)}_{\mathrm{in}}(x)
  :=
  \sum_{i=1}^N \frac{1}{2mc^2}
  \Big(\widehat E_i - A_i(x;\bm t)\Big)^{2},
  \label{eq:H-in-x}
\end{equation}
so that the associated interacting CS equation reads
\begin{equation}
  i\hbar c\,\partial_x\Psi(x,\bm t)
  =
  \mathcal H^{(N)}_{\mathrm{in}}(x)\,\Psi(x,\bm t).
\end{equation}
Here the $A_i(x;\bm t)$ are real functions encoding one-body and
many-body couplings in the temporal configuration coordinates. We
take them of the general form
\begin{equation}
  A_i(x;\bm t)
  =
  U(x,t_i)
  + \frac{1}{2}\sum_{j\ne i}\,\Omega\big(x;\,t_i,t_j\big),
  \label{eq:A-i-gen}
\end{equation}
with a one-body Carroll scalar $U(x,t)$ and a symmetric two-time
kernel $\Omega(x;t,t')=\Omega(x;t',t)$.

A particularly transparent case is when the $A_i$ arise from a single
scalar phase $S(x;\bm t)$, i.e.
\begin{equation}
  A_i(x;\bm t)=\partial_{t_i}S(x;\bm t),
  \qquad i=1,\ldots,N.
\end{equation}
Defining
\begin{equation}
  \Psi(x,\bm t)
  := \exp\!\Big(\frac{i}{\hbar}S(x;\bm t)\Big)\,\Phi(x,\bm t),
  \label{eq:Psi-from-Phi-S}
\end{equation}
one readily checks that $\Phi$ satisfies an
\emph{outside-the-square} CS equation
\begin{equation}
  i\hbar c\,\partial_x\Phi(x,\bm t)
  =
  \bigg[
     \sum_{i=1}^{N}\frac{\widehat E_i^{\,2}}{2mc^2}
     + V_t(\bm t)
  \bigg]\Phi(x,\bm t),
  \label{eq:CS-temporal-Sch}
\end{equation}
provided $S$ is chosen as
\begin{equation}
  S(x;\bm t) := \frac{x}{c}\,V_t(\bm t).
  \label{eq:S-def}
\end{equation}
In that case
\begin{equation}
  A_i(x;\bm t)
  = \partial_{t_i}S(x;\bm t)
  = \frac{x}{c}\,\partial_{t_i}V_t(\bm t),
  \qquad i=1,\ldots,N,
  \label{eq:A-general-map}
\end{equation}
and the temporal potential $V_t(\bm t)$ can be decomposed as
\begin{equation}
  V_t(\bm t)
  = \sum_{i=1}^{N}U_t(t_i)
    +\sum_{1\le i<j\le N}W_t(t_i,t_j),
  \label{eq:Vt-general}
\end{equation}
with one-body contributions $U_t$ and symmetric pair potentials
$W_t$. In other words, any choice of $V_t(\bm t)$ defines, via
\eqref{eq:S-def}–\eqref{eq:A-general-map}, an inside-the-square model
\eqref{eq:H-in-x} that is unitarily equivalent to the outside-the-square
form \eqref{eq:CS-temporal-Sch}. The converse is more restrictive:
generic couplings of the form \eqref{eq:A-i-gen} need not admit such a
scalar $S$ and may not correspond to a simple local $V_t(\bm t)$.

\subsection{Temporal oscillators}

As a concrete illustration, consider the temporal analogue of coupled harmonic oscillators. The spatial potential energy for $N$ oscillators on a line is given by
\[
  V_{\mathrm{osc}}(\mathbf x)
  = \sum_{i=1}^{N}\frac{1}{2}m\omega^{2}x_i^{2}
    +\frac{k_c}{2}\sum_{i<j}(x_i-x_j)^{2}.
\]
Formally replacing $x_i\!\to t_i$ and introducing temporal parameters $\omega_t$ and $k_t$, we define the temporal potential
\begin{equation}
  V_{t}^{\mathrm{osc}}(\bm t)
  = \sum_{i=1}^{N}\frac{1}{2}m\omega_{t}^{2}t_i^{2}
    +\frac{k_t}{2}\sum_{i<j}(t_i-t_j)^{2}.
  \label{eq:Vt-osc}
\end{equation}
It is important to note that $\omega_t$ in this temporal viewpoint is not a standard temporal frequency, as time here acts as the configuration coordinate. Instead, since $x$ is the evolution parameter, $\omega_t$ plays a role analogous to a scaled spatial wavenumber that determines the rate of oscillation as the system evolves along $x$, with units of acceleration. The corresponding outside-the-square CS equation is
\begin{equation}
  i\hbar c\,\partial_x\Phi(x,\bm t)
  =
  \sum_{i=1}^{N}\frac{\widehat E_i^{\,2}}{2mc^2}\,\Phi(x,\bm t)
  + V_{t}^{\mathrm{osc}}(\bm t)\,\Phi(x,\bm t),
  \label{eq:CS-osc-outside}
\end{equation}
which describes an $N$-body system of \emph{temporal oscillators}. On each equal-$x$ slice, the temporal coordinates $t_i$ exhibit the same normal-mode structure as the spatial coordinates $x_i$ do in the usual coupled-oscillator model.

To find the equivalent inside-the-square representation, we choose the scalar phase $S(x;\bm t)=\frac{x}{c}V_t^{\mathrm{osc}}(\bm t)$. The associated gauge couplings \eqref{eq:A-general-map} become
\begin{equation}
  A_i^{\mathrm{osc}}(x;\bm t)
  = \frac{x}{c}\,\partial_{t_i}V_{t}^{\mathrm{osc}}(\bm t)
  = \frac{x}{c}\left[
      \big(m\omega_t^2 + N k_t\big)\,t_i
      - k_t\sum_{j=1}^{N} t_j
    \right],
  \label{eq:A-osc}
\end{equation}
which defines a Hamiltonian of the form \eqref{eq:H-in-x} that is unitarily equivalent to \eqref{eq:CS-osc-outside}.

Since the generator in \eqref{eq:CS-osc-outside} is quadratic in both the temporal coordinates and the temporal momentum operators $\widehat E_i$, it can be diagonalized algebraically. Letting $\tau_\alpha$ be the temporal normal coordinates with corresponding effective frequencies $\omega_\alpha$, we define the temporal annihilation and creation operators
\begin{equation}
  \hat{b}_\alpha
  = \sqrt{\frac{mc^2 (\omega_\alpha/c)}{2\hbar}}
    \left( \tau_\alpha + \frac{i}{mc^2 (\omega_\alpha/c)} \widehat E_{\alpha} \right),
  \quad
  \hat{b}_\alpha^\dagger
  = \sqrt{\frac{mc\omega_\alpha}{2\hbar}}
    \left( \tau_\alpha - \frac{i}{mc\omega_\alpha} \widehat E_{\alpha} \right),
\end{equation}
where $\widehat E_\alpha = -i\hbar\partial_{\tau_\alpha}$. These operators satisfy $[\hat{b}_\alpha, \hat{b}_\beta^\dagger] = \delta_{\alpha\beta}$. The many-body Carroll generator then takes the diagonal form
\begin{equation}
  H_{total}
  = \sum_{\alpha=1}^N \hbar \left(\frac{\omega_\alpha}{c}\right)
    \left( \hat{b}_\alpha^\dagger \hat{b}_\alpha + \frac{1}{2} \right).
\end{equation}
The eigenvalues $\hbar\omega_\alpha/c$ correspond to the discrete wavenumbers for spatial propagation along $x$.

Physically, this model describes a mechanism for \emph{temporal synchronization}. Just as a spatial spring enforces a fixed distance between particles, the pairwise coupling $k_t(t_i - t_j)^2$ in \eqref{eq:Vt-osc} imposes an energetic penalty on asynchrony. A strong coupling $k_t$ effectively creates a ``rigid body in time,'' forcing the particles to arrive at any spatial slice $x$ with fixed time delays relative to one another. In the presence of this potential, the $N$-body wave packet does not disperse freely along the temporal axes; instead, the interaction acts as a synchronizer, protecting the temporal coherence of the packet and ensuring that multiparticle correlations (such as the HBT signals discussed in Sec.~\ref{sec:time-exchange-HBT}) are maintained over long propagation distances.

\subsubsection*{Minimal coupling description (Inside-the-square)}

Alternatively, we may describe the system via the inside-the-square representation by identifying the gauge couplings $A_i$ directly with the temporal potential energy function. For the $N=2$ case, we use the symmetric gauge choice
\begin{align}
  A_1(t_1, t_2) &= \frac{1}{2}m\omega_t^2 t_1^2 + \frac{k_t}{4}(t_1 - t_2)^2, \\[0.5em]
  A_2(t_1, t_2) &= \frac{1}{2}m\omega_t^2 t_2^2 + \frac{k_t}{4}(t_2 - t_1)^2.
\end{align}
The Hamiltonian is defined as
\begin{equation}
  \mathcal{H}_{\mathrm{in}}^{(2)} = \frac{1}{2mc^2} \left[ (\widehat E_1 - A_1)^2 + (\widehat E_2 - A_2)^2 \right].
\end{equation}
It is important to note that this inside-the-square construction is not generated by a global scalar phase $S$ and is therefore not equivalent to the outside-the-square form (2.12). Instead, it is obtained by directly coupling an oscillator-like potential to the energy operator in the Carroll--Schr\"odinger equation. Since this Hamiltonian is independent of the evolution coordinate $x$, we seek stationary solutions of the form $\Psi(x, \bm t) = e^{-i\mathcal{E}x/\hbar c}\Psi(\bm t)$, satisfying the eigenvalue equation $\mathcal{H}_{\mathrm{in}}^{(2)}\Psi(\bm t) = \mathcal{E}\Psi(\bm t)$.

To obtain the general solution, we separate the system into center-of-time $T=(t_1+t_2)/2$ and relative time $\tau=t_1-t_2$ coordinates. Substituting the potentials into the Hamiltonian yields
\begin{equation}
  \mathcal{H}_{\mathrm{in}}^{(2)} = \frac{1}{4mc^2} \Big(\widehat E_T - A_\Sigma(T, \tau)\Big)^2 + \frac{1}{mc^2} \Big(\widehat E_\tau - A_\Delta(T, \tau)\Big)^2,
\end{equation}
where the sum and difference potentials are
\begin{equation}
  A_\Sigma = m\omega_t^2 T^2 + \left(\frac{1}{4}m\omega_t^2 + \frac{k_t}{2}\right)\tau^2,
  \qquad
  A_\Delta = \frac{1}{2}m\omega_t^2 T \tau.
\end{equation}

We perform a unitary gauge transformation to decouple the interaction term $A_\Delta$. Introducing the phase $\Lambda(T, \tau) = \frac{1}{4}m\omega_t^2 T \tau^2$, we transform the wavefunction as $\Psi = e^{i\Lambda/\hbar} \tilde{\Psi}$. This shifts the momentum operators as $\widehat E_\tau \to \widehat E_\tau + A_\Delta$ (canceling the coupling in the second term) and $\widehat E_T \to \widehat E_T + \frac{1}{4}m\omega_t^2 \tau^2$. The transformed Hamiltonian is separable
\begin{equation}
  \tilde{\mathcal{H}} = \frac{1}{4mc^2} \left( \widehat E_T - m\omega_t^2 T^2 - \frac{k_t}{2}\tau^2 \right)^2 + \frac{1}{mc^2} \widehat E_\tau^2.
\end{equation}

We define the operator $\hat{\mathcal{O}}_T \equiv \widehat E_T - m\omega_t^2 T^2$. Since $\hat{\mathcal{O}}_T$ acts only on $T$ and $\widehat E_\tau^2$ acts only on $\tau$, these two operators commute. We can therefore construct the general solution as a product of their simultaneous eigenstates, $\tilde{\Psi}(T, \tau) = f_\lambda(T) \phi_\lambda(\tau)$. The function $f_\lambda(T)$ satisfies the first-order differential equation $\hat{\mathcal{O}}_T f_\lambda = \lambda f_\lambda$, with solution
\begin{equation}
  f_\lambda(T) = \exp\left[ \frac{i}{\hbar} \left( \frac{1}{3}m\omega_t^2 T^3 + \lambda T \right) \right].
\end{equation}
Substituting the eigenvalue $\lambda$ back into the Hamiltonian equation for $\phi_\lambda(\tau)$, and using the explicit differential form $\widehat E_\tau = -i\hbar\partial_\tau$, we obtain the second-order differential equation for the relative motion
\begin{equation}
  -\frac{\hbar^2}{mc^2} \frac{d^2 \phi_\lambda}{d\tau^2} + \left[ \frac{1}{4mc^2} \left( \lambda - \frac{k_t}{2}\tau^2 \right)^2 \right] \phi_\lambda(\tau) = \mathcal{E} \phi_\lambda(\tau).
\end{equation}
This effective potential is a quartic well for any real $\lambda$, leading to a discrete energy spectrum and bound states. Thus, the general solution confirms that the interaction $k_t$ enforces synchronization through temporal confinement, modulated by the continuous parameter $\lambda$. It is important to note that this inside-the-square construction is not generated by a global scalar phase $S$, and therefore it is not  equivalent to the outside-the-square solution from the previous subsection. Rather, it arises by directly coupling an oscillator-like potential to the energy operator in our Carroll Schrödinger equation.

\section{Spatial interactions viewpoint and model systems}
\label{sec:space-viewpoint}

An important question is how a multi-particle Carroll--Schr\"odinger (CS) system responds to \emph{purely spatial} drives, i.e.\ when the interaction energy $U(\mathbf x)$ depends only on the spatial configuration, as in standard classical and Schr\"odinger dynamics. Physically, this description corresponds to the Carrollian limit of a multiparticle Klein--Gordon system coupled to static spatial potentials; for internal interactions, such potentials are typically defined in the center-of-mass frame where the interaction energy acts instantaneously~\cite{Horwitz1996}. After examining solution behavior, continuity laws and operator structure, it is natural in $1{+}1$ dimensions to regard the CS dynamics as an \emph{evolution in $x$} with $t$ as the configuration variable \cite{Rojas2025}. In order to compare spatial behavior on a more equal footing with standard Schr\"odinger systems, in this section we adopt a complementary viewpoint where we compare the usual spatial probability density $|\psi_{\rm Sch}|^{2}$ with the Carroll \emph{temporal} probability current $J_t$, which from a purely spatial point of view can be interpreted as an effective spatial density. It must be emphasized that in the strict Carrollian limit ($c \to 0$) space becomes absolute and causally disconnected (``frozen'') \cite{deBoer2022}, meaning the physical degrees of freedom reside entirely in the temporal domain; however, we use this spatial viewpoint as a formal tool to illustrate how external potentials would act when time is treated as the evolution parameter.

The CS continuity equation (in the free case) can be written as
\begin{equation}
\partial_x \rho_t(x,t)+\partial_t J_t(x,t)=0,
\qquad
\rho_t(x,t):=|\Psi(x,t)|^{2},
\qquad
J_t(x,t):=\frac{\hbar}{mc^{3}}\operatorname{Im}\!\big[\Psi^*\partial_t\Psi\big],
\label{eq:CS-cont}
\end{equation}
where the subscript $t$ emphasizes that $\rho_t$ and $J_t$ are the density and
current associated with the time coordinate. Exchanging the roles of density
and current, the same equation can be written as
\begin{equation}
\partial_x J_x(x,t)+\partial_t \rho_x(x,t)=0,
\qquad
\rho_x(x,t):=\frac{\hbar}{mc^{3}}\operatorname{Im}\!\big[\Psi^*\partial_t\Psi\big],
\qquad
J_x(x,t):=|\Psi(x,t)|^{2},
\label{eq:CS-cont2}
\end{equation}
so that, in $1{+}1$ dimensions, the dynamics can be viewed either as an
$x$-evolution with $\rho_t$ a positive-definite norm, or as a $t$-evolution
with $\rho_x$ playing the role of an effective spatial density (note that
$\rho_x$ is not positive definite, so it is not a genuine probability density,
only an effective density from the dual viewpoint). In the present work, the
first viewpoint is preferred (except in this section), since it provides a
unitary evolution in $x$ with a positive definite probability density $\rho_t$.

For these reasons, in this section we depart from the $N$-time configuration space of Sec.~2 and consider a more standard setting with $N$ spatial coordinates $\mathbf x$ and a single time $t$. From this spatial viewpoint we regard $t$ as the evolution parameter and $\mathbf x=(x_1,\dots,x_N)$ as the configuration coordinates, so that the wave function is $\Psi(\mathbf x,t)$ rather than $\Psi(x;\,t_1,\dots,t_N)$. This is done solely to compare against familiar many-body Schr\"odinger systems in a compact way; it is not the same $N$-time configuration used in Sec.~2. For a configuration $\mathbf x=(x_1,\ldots,x_N)$ and a time-independent total potential energy $U_{\!tot}(\mathbf x)$, we start from the CS evolution written in the inside-the-square form

\begin{equation}
  i\hbar c\,\sum_{j=1}^{N}\partial_{x_j}\Psi(\mathbf x,t)\;=\;
  \frac{1}{2mc^{2}}\Big(-i\hbar\,\partial_t - U_{\!tot}(\mathbf x)\Big)^{2}\Psi(\mathbf x,t),
  \label{eq:CS-inside}
\end{equation}
and define the gauge–transformed field
\begin{equation}
  \Phi(\mathbf x,t)\;=\;
  \exp\!\Big(-\frac{i}{\hbar}\!\int^{t}\!U_{\!tot}(\mathbf x)\,d\tau\Big)\,\Psi(\mathbf x,t)
  =\exp\!\Big(-\frac{i}{\hbar}(t-t_0)U_{\!tot}(\mathbf x)\Big)\,\Psi(\mathbf x,t),
  \label{eq:gauge-def}
\end{equation}
where $t_0$ is a reference time. Substituting into \eqref{eq:CS-inside} and dividing out the common phase gives
\begin{equation}
  i\hbar c\,\sum_{j=1}^{N}\partial_{x_j}\Phi(\mathbf x,t)
  \;=\;-\,\frac{\hbar^{2}}{2mc^{2}}\;\partial_t^{2}\Phi(\mathbf x,t)
        \;+\;c\,\sum_{j=1}^{N}\mathcal F_j(\mathbf x,t)\,\Phi(\mathbf x,t),
  \quad
  \mathcal F_j(\mathbf x,t):=\int^{t}\!\partial_{x_j}U_{\!tot}(\mathbf x)\,d\tau.
  \label{eq:CS-outside}
\end{equation}
Because $U_{\!tot}$ is independent of $t$, we have
$\mathcal F_j(\mathbf x,t)=(t-t_0)\,\partial_{x_j}U_{\!tot}(\mathbf x)$, and the
equation simplifies to
\begin{equation}
 i\hbar c\,\sum_{j=1}^{N}\partial_{x_j}\Phi(\mathbf x,t)
  \;=\;-\,\frac{\hbar^{2}}{2mc^{2}}\;\partial_t^{2}\Phi(\mathbf x,t)
        \;+\;c\,(t-t_{0})\,\sum_{j=1}^{N}\partial_{x_j}U_{\!tot}(\mathbf x)\,\Phi(\mathbf x,t).
  \label{eq:CS-outside-space-only}
\end{equation}
Even for a static $U_{\!tot}(\mathbf x)$ the effective drive in this
linearized picture depends on both $\mathbf x$ and $t$ through the factor
$c\,(t-t_0)\,\partial_{x_j}U_{\!tot}(\mathbf x)$, so the dynamics is generally
nonseparable and solutions are not stationary in $t$.

For later comparison it is useful to recall that, for the same
$U_{\!tot}(\mathbf x)$, the standard multi–particle Schr\"odinger evolution is
\begin{equation}
  i\hbar\,\partial_t \psi(\mathbf x,t)
  = -\,\frac{\hbar^{2}}{2m}\sum_{j=1}^{N}\partial_{x_j}^{2}\psi(\mathbf x,t)
    + U_{\!tot}(\mathbf x)\,\psi(\mathbf x,t),
  \label{eq:Sch-eq}
\end{equation}
with $|\psi(\mathbf x,t)|^{2}$ the spatial probability density.

\subsection{Model case 1: Coupled quantum harmonic oscillators}
\label{subsec:spatial-oscillators}

As a first example of multi–particle Carrollian dynamics with spatial
interactions, consider two coupled harmonic oscillators with total potential
energy
\begin{equation}
U_{\!tot}(x_1,x_2)
= \frac{1}{2}m\omega^2 (x_{1}^2 + x_{2}^2 )
  + \frac{1}{2}k_{c} (x_{1} - x_{2})^2 .
\label{eq:Utot-2osc}
\end{equation}
Substituting into \eqref{eq:CS-outside-space-only} we require
$\sum_j\partial_{x_j}U_{\!tot}$, which yields
\begin{equation}
\sum_{j=1}^{2}\partial_{x_j}U_{\!tot}(x_1,x_2)
= m\omega^{2}(x_1+x_2).
\label{eq:dUtot-2osc}
\end{equation}
Equivalently,
\begin{equation}
\sum_{j=1}^{2}\partial_{x_j}\big[c(t-t_0)U_{\!tot}(x_1,x_2)\big]
= m\omega^{2}c(t-t_0)(x_1+x_2),
\label{eq:dUtot-2osc-ct}
\end{equation}
The corresponding evolution equation reads
\begin{equation}
  i\hbar c\,(\partial_{x_1}+\partial_{x_2})\Phi(x_1,x_2,t)
  =-\frac{\hbar^{2}}{2mc^{2}}\partial_t^{2}\Phi(x_1,x_2,t)
  +c(t-t_{0})\,m\omega^{2}(x_1+x_2)\Phi(x_1,x_2,t),
  \label{eq:CS-2body-pre}
\end{equation}
and introducing collective and relative coordinates
\begin{equation}
U:=x_1+x_2,\qquad V:=x_1-x_2,
\label{eq:UV-def-2osc}
\end{equation}
we have \(\partial_{x_1}+\partial_{x_2}=2\,\partial_U\). Defining \(x:=U/2\)
removes this factor and leads to the one–dimensional evolution
\begin{equation}
  i\hbar c\,\partial_x\Phi(x,V,t)
  =-\frac{\hbar^{2}}{2mc^{2}}\partial_t^{2}\Phi(x,V,t)
  +c(t-t_{0})\,k_{\!{\rm eff}}\,x\,\Phi(x,V,t),
  \qquad k_{\!{\rm eff}}:=2m\omega^{2}.
  \label{eq:CS-2body-1D}
\end{equation}
Multiplying by \(2mc^{2}\) gives
\begin{equation}
  i\hbar(2mc^{3})\partial_x\Phi
  =-\hbar^{2}\partial_t^{2}\Phi
  + (2mc^{3})\,k_{\!{\rm eff}}x(t-t_0)\Phi,
  \label{eq:CS-2body-1D-multiplied}
\end{equation}

Solving \eqref{eq:CS-2body-1D} via Fourier transform and the method of characteristics, we find that the resulting kernel is

\begin{equation}
\begin{aligned}
K^{(2)}(U,V;\,t,t')
  &= \sqrt{\frac{mc^{3}}{\pi i\hbar\,U}}\,
     \exp\Bigg[
       -\frac{i m c^{3}}{\hbar U}
       \Big(
         (t{-}t_{0})-(t'{-}t_{0})
         -\frac{k_{\!{\rm eff}}U^{3}}{24 m c^{3}}
         -\frac{C(V)U^{2}}{3c}
       \Big)^{2} \\
  &\hspace{4em}
       -\frac{i\,k_{\!{\rm eff}}^{2}U^{5}}{320\,\hbar\,m c^{3}}
       +\frac{i\,k_{\!{\rm eff}}U^{2}}{4\hbar}(t{-}t_{0})
     \Bigg].
\end{aligned}
\label{eq:K2-corrected}
\end{equation}
with
\begin{equation}
  C(V)=\frac{1}{\hbar}
  \left(\frac{m\omega^{2}}{4}+\frac{k_{c}}{2}\right)V^{2}.
  \label{eq:CV-def}
\end{equation}
The additional term proportional to \(C(V)\) encodes the coupling through the
integration constants of the characteristic equations.

The general solution with boundary data \(\Phi_0(V,t)=\Phi(U=0,V,t)\) is then
\begin{equation}
  \Phi(x_1,x_2,t)
  =\int_{-\infty}^{\infty}\!dt'\;
  K^{(2)}(U,V;\,t,t')\,\Phi_0(V,t'),
  \qquad U:=x_1+x_2,\;V:=x_1-x_2.
  \label{eq:Phi-2body-prop-corrected}
\end{equation}
For illustrative purposes we will choose a Gaussian initial profile
\begin{equation}
  \Phi_0(V,t)
  =f(V)\,\frac{1}{(2\pi\sigma^2)^{1/4}}
  \exp\!\left[-\frac{(t-t_0)^{2}}{4\sigma^{2}}\right],
  \label{eq:Phi0-Gauss-def}
\end{equation}
the integral remains Gaussian and yields
\begin{equation}
\begin{aligned}
\Phi(x_1,x_2,t)
&= f(V)\,\frac{1}{(2\pi)^{1/4}\sqrt{\Sigma(U)}}
   \exp\!\left[-\frac{\big((t{-}t_{0})-t_c(U,V)\big)^{2}}{4\Sigma^{2}(U)}\right] \\
&\quad\times
   \exp\!\left\{\,i\Big[\chi(U)\big((t{-}t_{0})-t_c(U,V)\big)^{2}
   +\frac{k_{\!{\rm eff}} U^{2}}{4\hbar}(t{-}t_{0})
   +\varphi(U,V)\Big]\right\},
\end{aligned}
\label{eq:Phi-2body-Gauss-corrected}
\end{equation}
where
\begin{equation}
  \Sigma^{2}(U)=\sigma^{2}+\frac{\hbar^{2}U^{2}}{16\sigma^{2}m^2 c^6},\qquad
  t_c(U,V)=\frac{k_{\!{\rm eff}}U^{3}}{24 m c^{3}}
  +\frac{C(V)U^{2}}{3c},\qquad
  \chi(U)=\frac{\hbar U}{32\sigma^{2} m c^3 \,\Sigma^{2}(U)}.
  \label{eq:Sigma-tc-chi-V}
\end{equation}
The global phase can be chosen as
\begin{equation}
\varphi(U,V)
=-\frac{k_{\!{\rm eff}}^{2}U^{5}}{320\,\hbar\,m c^{3}}
-\frac12\arctan\!\Big(\frac{\hbar U}{2\sigma^{2}c}\Big)
-\frac{C(V)^{2}U}{2mc^{3}},
\label{eq:varphi-UV-def}
\end{equation}
so that the normalization remains real and continuous at \(U=0\).

Returning to the original field,
\begin{equation}
\Psi(x_1,x_2,t)
=\exp\!\Big(\tfrac{i}{\hbar}(t-t_0)\,U_{\!tot}(x_1,x_2)\Big)
\,\Phi(x_1,x_2,t),
\qquad |\Psi|^{2}=|\Phi|^{2},
\label{eq:Psi-from-Phi-corrected}
\end{equation}
we obtain the probability density and temporal current,
\begin{equation}
\rho_{t}^{(2)}(x_1,x_2,t)
=\frac{|f(x_1 - x_2)|^{2}}{\sqrt{2\pi}\,\Sigma(x_1 + x_2)}\,
\exp\!\left[-\frac{\big((t{-}t_{0})-t_c(x_1 + x_2,\,x_1-x_2)\big)^{2}}{2\Sigma^{2}(x_1 + x_2)}\right],
\label{eq:rhoS-2-Psi-corrected}
\end{equation}
\begin{equation}
J^{(2)}_{t}(x_1,x_2,t)
=\rho_{t}^{(2)}(x_1,x_2,t)\left[
\frac{\omega^{2}(x_1 + x_2)^{2}}{2c^{3}}
+\frac{C(x_1 - x_2)}{c^{2}}
+\frac{\hbar^{2}(x_1 + x_2)}{4m c^{4}\sigma^{2}\Sigma^{2}(x_1 + x_2)}
\big((t{-}t_{0})-t_c(U,\,V)\big)
\right],
\label{eq:J-2-closed-corrected}
\end{equation}
describing a temporally Gaussian Carroll fluid whose width and chirp depend on
the collective coordinate \(U=x_1+x_2\), while the coupling–induced drift
$t_c(U,V)$ introduces a controlled temporal asymmetry along the relative
coordinate.

\clearpage

\begin{figure*}[t]
  \centering
  \begin{subfigure}[b]{1.1\textwidth}
    \centering
    \includegraphics[width=\textwidth]{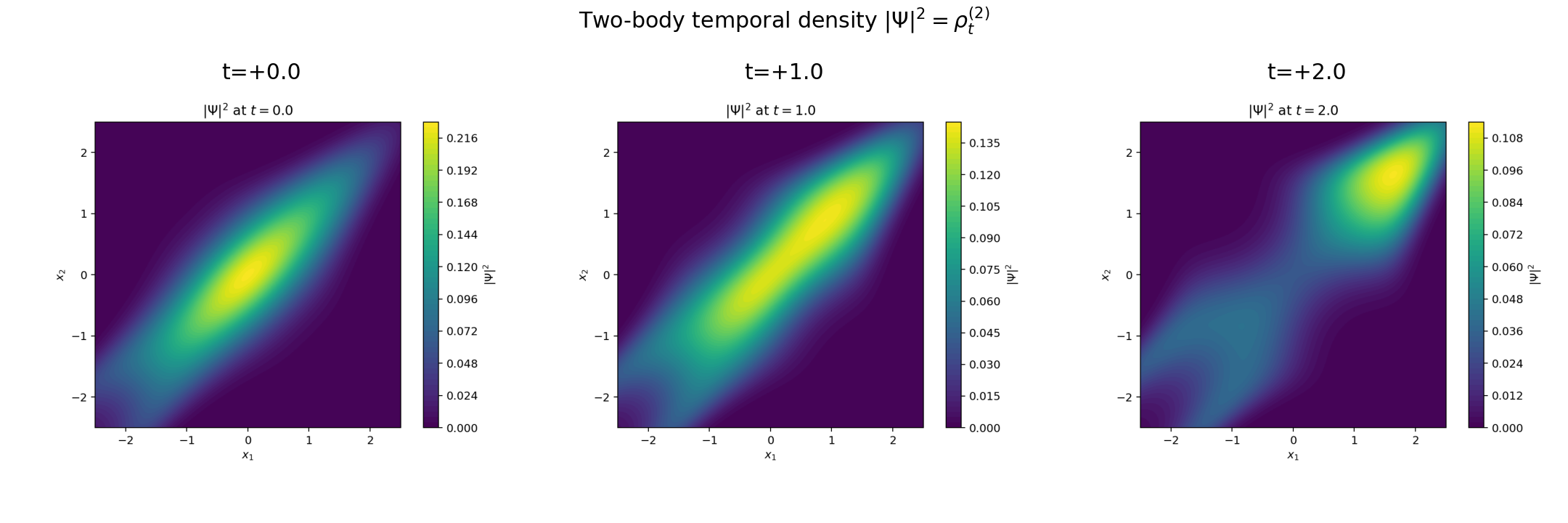}
    \caption{Two–body temporal density $\rho_t^{(2)}(x_1,x_2,t)=|\Psi|^2$ at three times.}
    \label{fig:rho2-triptych}
  \end{subfigure}\hfill
  \begin{subfigure}[b]{1.1\textwidth}
    \centering

    \includegraphics[width=\textwidth]{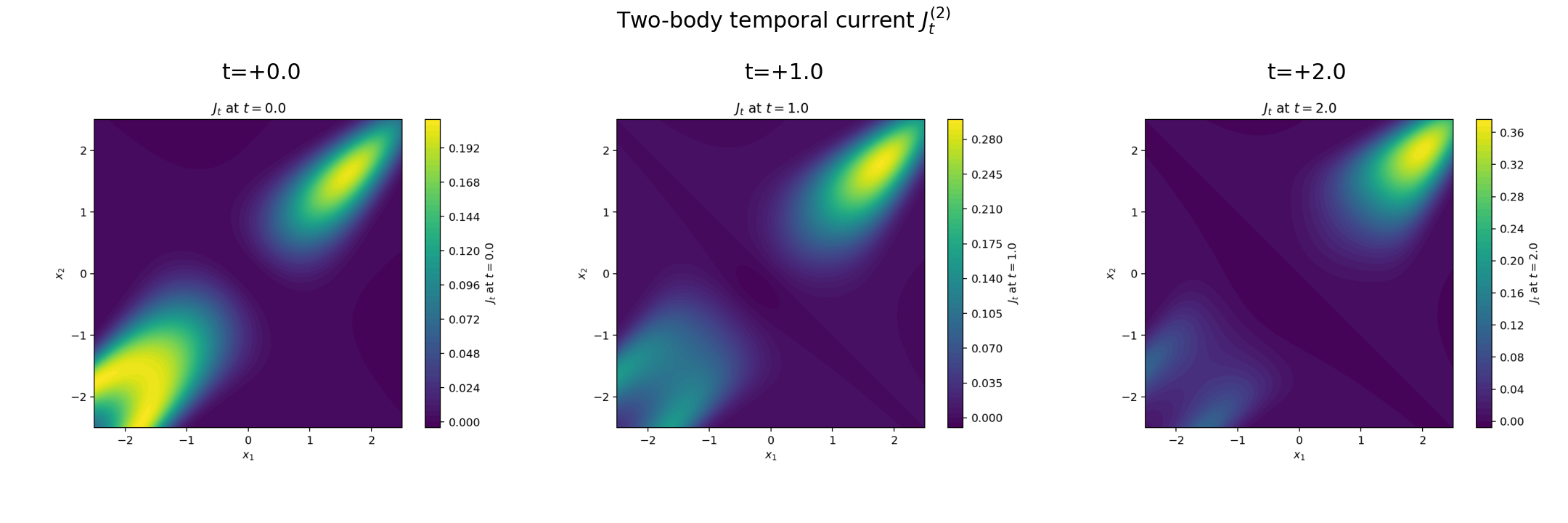}
    \caption{Two–body temporal current $J_t^{(2)}(x_1,x_2,t)$ (which can also be interpreted as an effective Spatial density $\rho_x^{(2)}(x_1,x_2,t)$ via Eq. \ref{eq:CS-cont2}) at three times.}
    \label{fig:J2-triptych}
  \end{subfigure}
  \caption{ Carroll evolution of the two–oscillator model in the
   $(x_1,x_2)$ plane.}
  \label{fig:two-body-density-current}
\end{figure*}

\begin{figure*}[t]
  \centering
  \begin{subfigure}[b]{0.5\textwidth}
    \centering
    \includegraphics[width=\textwidth]{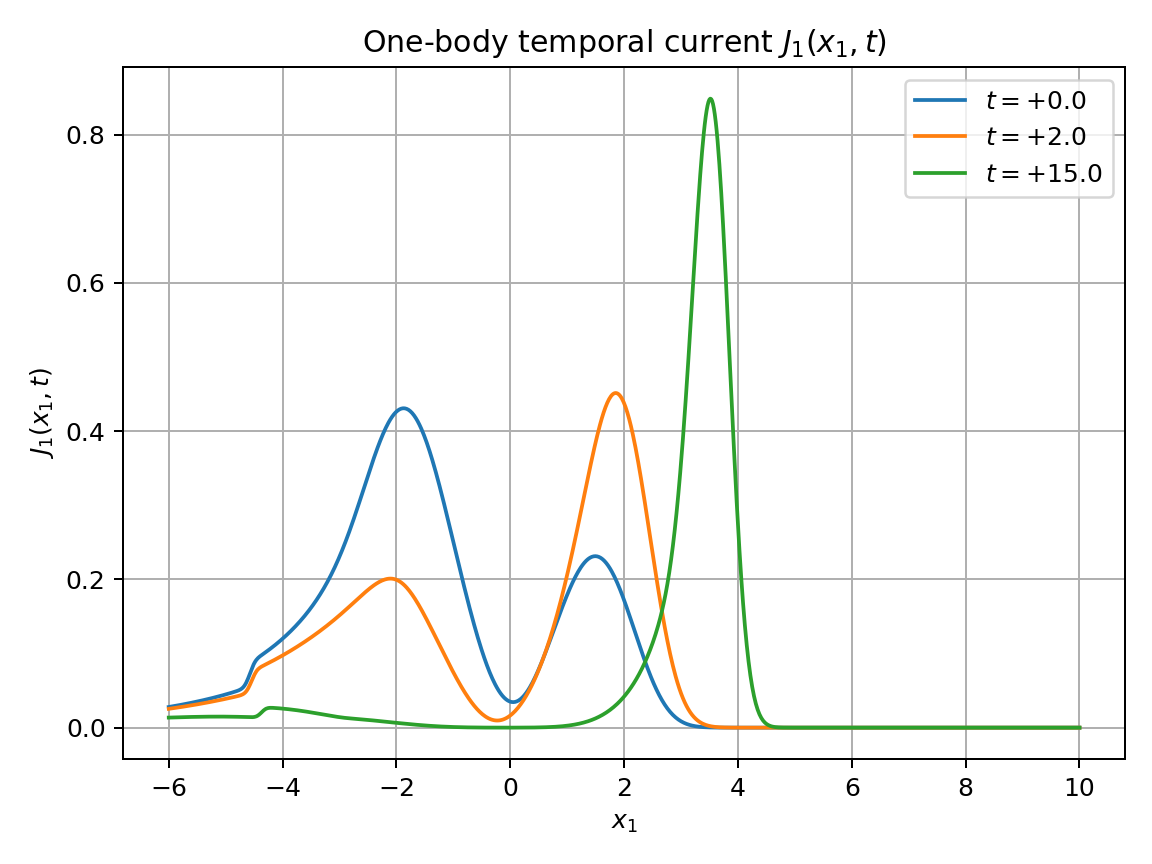}
    \caption{One–body temporal current $J_1(x_1,t)$ at several times $t$.}
    \label{fig:J1-vs-x}
  \end{subfigure}\hfill
  \begin{subfigure}[b]{0.5\textwidth}
    \centering
    \includegraphics[width=\textwidth]{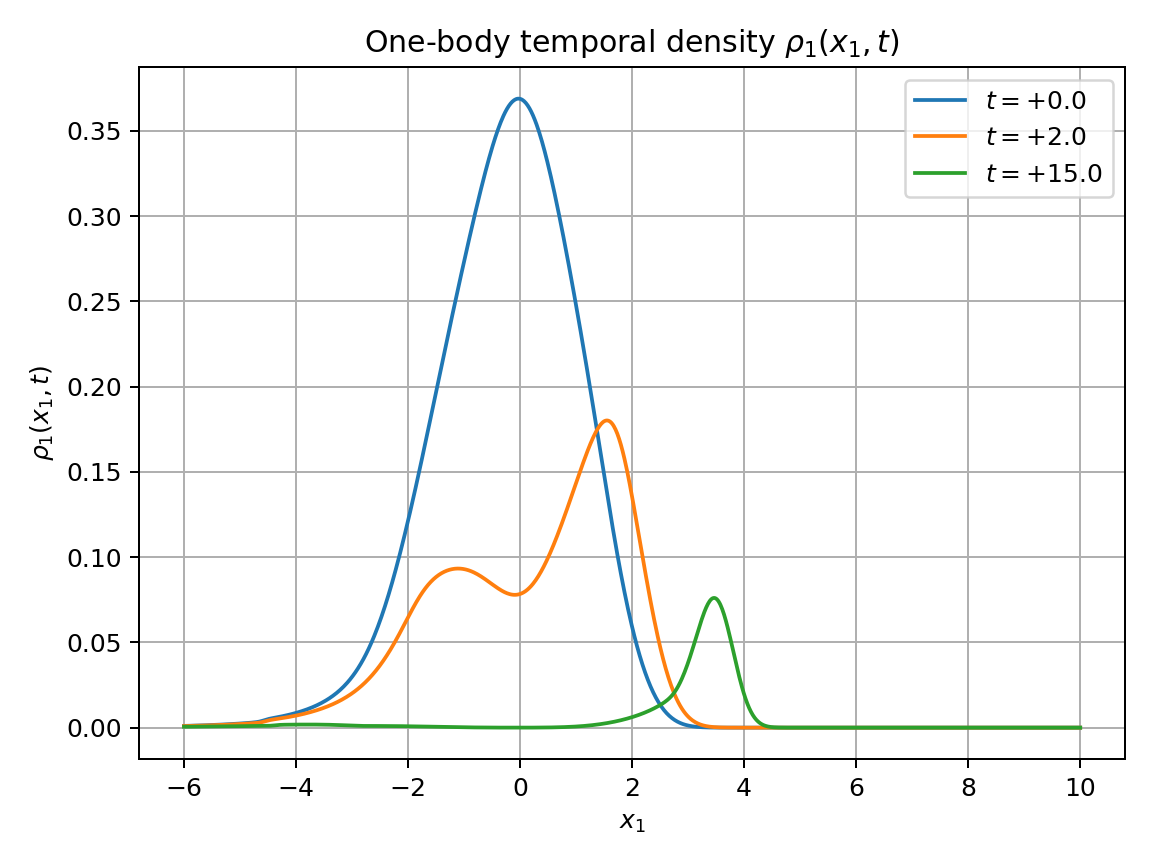}
    \caption{One–body temporal density $\rho_1(x_1,t)$ at the same times.}
    \label{fig:rho1-vs-x}
  \end{subfigure}
  \caption{Spatial profiles of the one–body observables at fixed times.}
  \label{fig:one-body-x}
\end{figure*}

\clearpage

\begin{figure*}[t]
  \centering
  \begin{subfigure}[b]{0.5\textwidth}
    \centering
    \includegraphics[width=\textwidth]{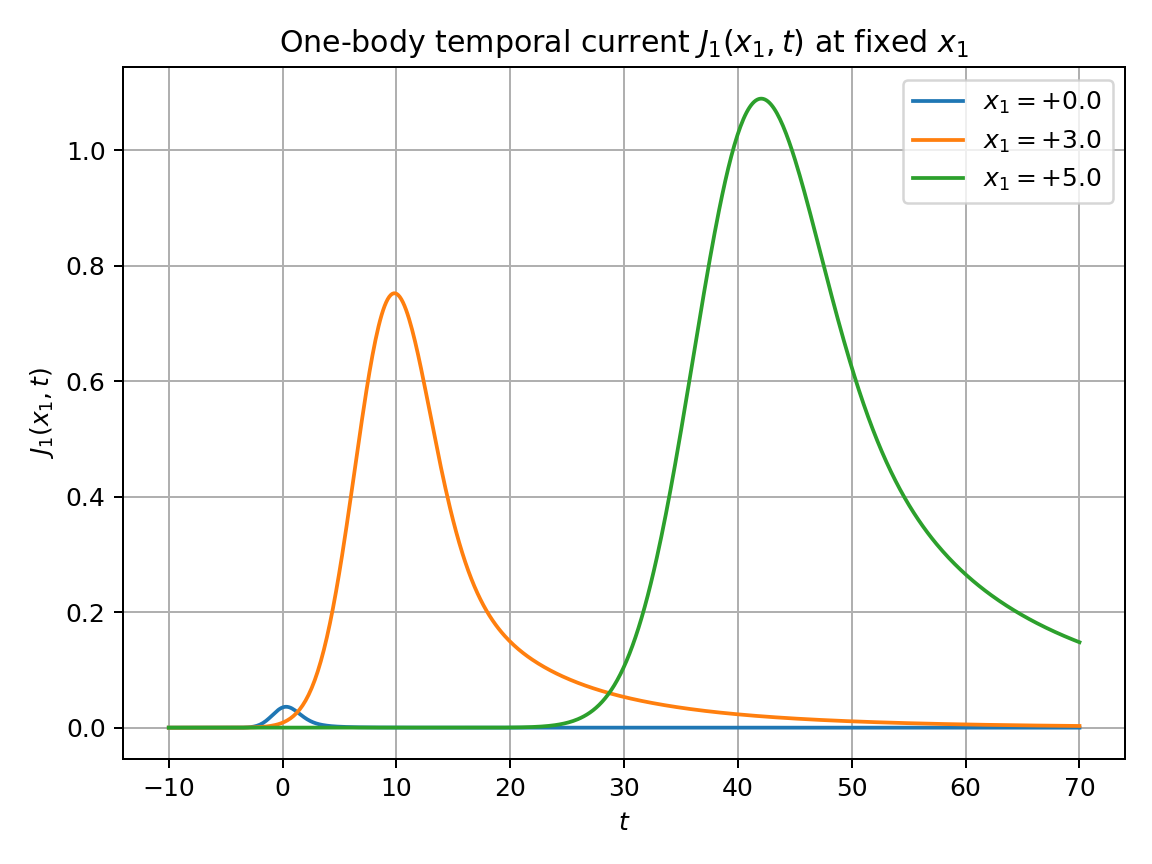}
    \caption{One–body temporal current $J_1(x_1,t)$ at several fixed positions $x_1$.}
    \label{fig:J1-vs-t}
  \end{subfigure}\hfill
  \begin{subfigure}[b]{0.5\textwidth}
    \centering
    \includegraphics[width=\textwidth]{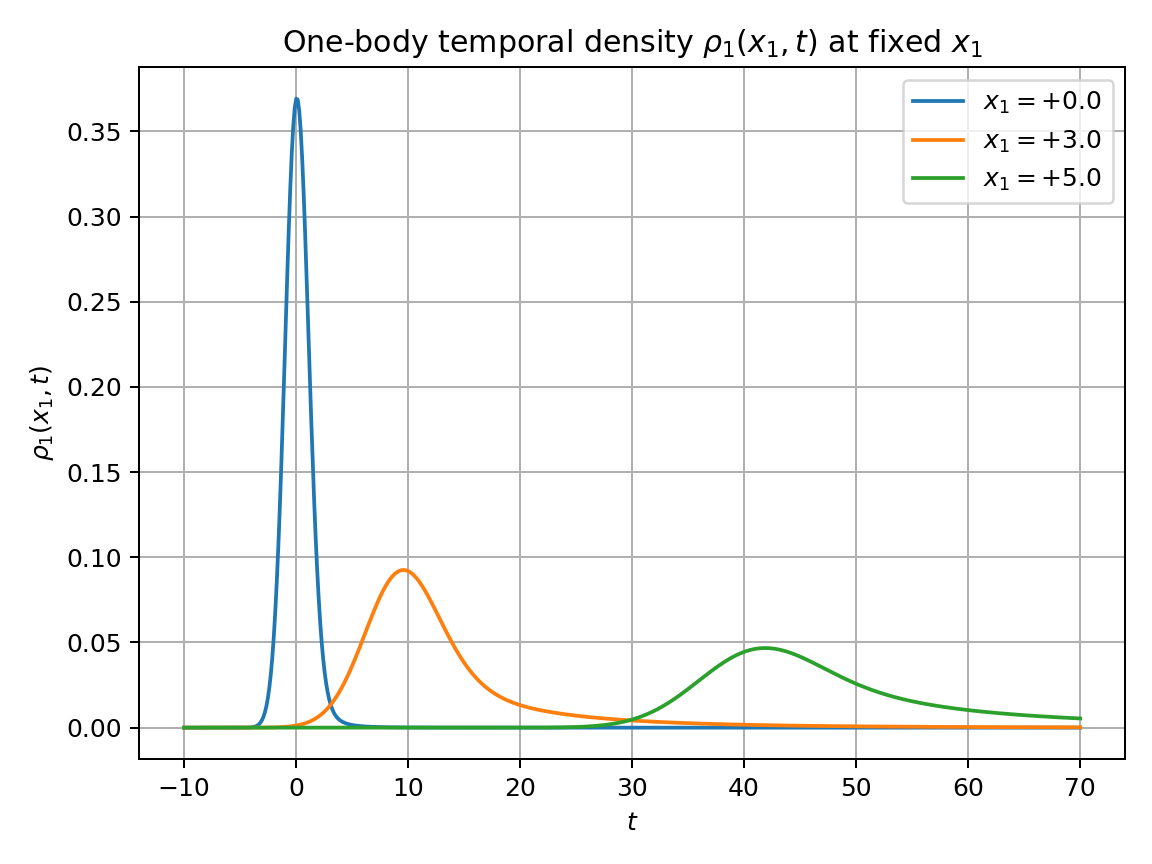}
    \caption{One–body temporal density $\rho_1(x_1,t)$ at the same positions.}
    \label{fig:rho1-vs-t}
  \end{subfigure}
  \caption{Temporal profiles of the one–body observables at fixed $x_1$.}
  \label{fig:one-body-t}
\end{figure*}

Figures~\ref{fig:two-body-density-current}–\ref{fig:one-body-t} illustrate the
dynamics of the explicit Gaussian solution
\eqref{eq:rhoS-2-Psi-corrected}–\eqref{eq:J-2-closed-corrected} for the
parameter choice $\hbar=c=m=k_c=1$, $\omega=0.7$, $\sigma=1$, $t_0=0$, and a
normalized, even Gaussian relative profile
$f(V)\propto\exp[-V^{2}/(2s_{\rm rel}^{2})]$ with $s_{\rm rel}=1$.  The Carroll
dynamics does not fix $f(V)$; we choose it from the spatial Schr\"odinger
partner (relative ground state or required symmetry) and then normalize it.
The genuinely Carrollian structure sits in the $(U,t)$ sector: the dependence
on the collective coordinate $U=x_1+x_2$ and on $t$ is governed by the kernel
\eqref{eq:K2-corrected} through the drift time $t_c(U,V)$, while $f(V)$ and
$C(V)$ control the detailed shape and parity along the relative direction.

In the two–body maps of Fig.~\ref{fig:two-body-density-current}, the density
$\rho_t^{(2)}$ forms an elongated packet concentrated near $x_1\simeq x_2$,
with a mild skew in the $V$–direction coming from the $C(V)$ term.  As $t$
increases the packet drifts towards larger $(x_1,x_2)$ and broadens along $U$,
consistent with the temporal Gaussian being centered near
$(t-t_0)\simeq t_c(U,V)$.  At fixed $t$ the probability peaks along the curve
in the $(x_1,x_2)$ plane where the drift time $t_c(U,V)$ matches the
observation time.  The current $J_t^{(2)}$ exhibits the same overall drift and
broadening, together with the increase in magnitude as the collective packet
moves to larger $U$, in agreement with the analytic prefactor multiplying
$\rho_t^{(2)}$ in \eqref{eq:J-2-closed-corrected} and with the additional
$C(V)/c^{2}$ contribution.

Figures~\ref{fig:one-body-x} and~\ref{fig:one-body-t} display the one–body observables obtained by integrating over $x_2$.  At fixed time (Fig.~\ref{fig:one-body-x}) the marginal density $\rho_1(x_1,t)$ and the current $J_1(x_1,t)$ develop single peaks that move to the right as $t$ grows and become sharper at later times, tracking the collective drift in $U$ and its temporal spreading.  Conversely, at fixed position (Fig.~\ref{fig:one-body-t}) each $x_1$ sees a well–defined temporal pulse: the signal switches on, reaches a maximum near the value of $t$ solving $t_c(U,V)\simeq t$ for the dominant values of $V$, and then decays as the packet moves past.  These plots make explicit that, in this purely spatial model, the CS evolution is \emph{slaved} to the boundary data on the plane $U=0$. Once the profile $\Phi_0(V,t')$ is fixed, the bulk dynamics is obtained essentially by convolution, meaning the spatial probability density is dominantly determined by the boundary profile $f(x_1-x_2)$. Unlike standard quantum mechanics, where interaction forces would dynamically reshape the relative wavefunction, here the evolution merely transports the initial relative configuration. This highlights the indifference of the bulk dynamics to the internal interaction potentials; the particles tend to maintain their initial relative spatial distribution rather than generating new correlations through dynamical evolution.

\subsection*{$N$–body generalization}

For $N$ coupled harmonic oscillators on a line with nearest–neighbour coupling,
the total potential energy is taken as
\begin{equation}
U_{\!tot}(\mathbf x)
=\frac{1}{2}m\omega^{2}\sum_{n=1}^{N}x_{n}^{2}
+\frac{1}{2}k_{c}\sum_{n=1}^{N-1}(x_{n+1}-x_{n})^{2},
\qquad
\mathbf x:=(x_1,\ldots,x_N).
\label{eq:Utot-N-chain}
\end{equation}
In the gauge–transformed outside–the–square equation
\eqref{eq:CS-outside-space-only}, $U_{\!tot}$ enters only through
$\sum_{j}\partial_{x_j}U_{\!tot}$. A short computation shows that all
$k_c$–dependent terms cancel pairwise and
\begin{equation}
\sum_{j=1}^{N}\partial_{x_j}U_{\!tot}(\mathbf x)
= m\omega^{2}\sum_{j=1}^{N}x_j.
\label{eq:dU-sum-N}
\end{equation}
Thus the coupling $k_c$ affects the dynamics only through the relative
coordinates, via the integration constants of the characteristic equations.

We introduce collective and relative coordinates
\begin{equation}
U:=\sum_{j=1}^{N}x_j,\qquad
\mathbf r:=(r_1,\ldots,r_{N-1}),\quad
r_a:=x_a-x_N\ (a=1,\ldots,N-1),
\qquad
x:=\frac{U}{N}.
\label{eq:collective-rel-N}
\end{equation}
This map $(U,\mathbf r)\leftrightarrow\mathbf x$ has constant Jacobian, spans
the hyperplane orthogonal to $(1,\ldots,1)$, and any other full–rank set of
$N{-}1$ differences is equivalent. Since
$\sum_{j}\partial_{x_j}=N\,\partial_{U}$ and $\partial_{U}=\tfrac1N\partial_x$,
we have $\sum_{j}\partial_{x_j}=\partial_x$, and the gauge–transformed field
$\Phi$ (defined by $\Psi=e^{\frac{i}{\hbar}(t-t_0)U_{\!tot}}\,\Phi$) obeys
\begin{equation}
i\hbar c\,\partial_x\Phi(x,\bm r,t)
=-\frac{\hbar^{2}}{2mc^{2}}\partial_t^{2}\Phi(x,\bm r,t)
+c(t-t_{0})\,k_{N}\,x\,\Phi(x,\bm r,t),
\qquad k_{N}:=N\,m\omega^{2}.
\label{eq:CS-N-1D}
\end{equation}
Multiplying by $2mc^{2}$ gives
\begin{equation}
i\hbar(2mc^{3})\,\partial_x\Phi
=-\hbar^{2}\partial_t^{2}\Phi
+(2mc^{3})\,k_{N}\,x(t-t_{0})\,\Phi.
\label{eq:CS-N-1D-c}
\end{equation}

It is convenient to separate the potential into collective and relative parts.
Using \eqref{eq:collective-rel-N} one finds
\begin{equation}
U_{\!tot}(U,\mathbf r)
=\frac{m\omega^{2}}{2N}\,U^{2}+U_{\rm rel}(\mathbf r),
\label{eq:Utot-split}
\end{equation}
with
\begin{equation}
U_{\rm rel}(\mathbf r)
=\frac{m\omega^{2}}{2}
\left[\sum_{a=1}^{N-1}r_a^{2}
-\frac{1}{N}\Big(\sum_{a=1}^{N-1}r_a\Big)^{2}\right]
+\frac{k_{c}}{2}\left[
\sum_{n=1}^{N-2}(r_{n+1}-r_{n})^{2}+r_{N-1}^{2}\right].
\label{eq:Urel-N}
\end{equation}
For later convenience we define the (dimensionful) coupling function
\begin{equation}
C(\mathbf r):=\frac{1}{\hbar}\,U_{\rm rel}(\mathbf r).
\label{eq:C-of-r-def}
\end{equation}
For $N=2$ one has $r_1=x_1-x_2=:V$ and \eqref{eq:Urel-N} reduces to
$U_{\rm rel}(V)=(\frac{m\omega^{2}}{4}+\frac{k_c}{2})V^{2}$, so that
\eqref{eq:C-of-r-def} reproduces the two–body expression
$C(V)=\big(\frac{m\omega^{2}}{4}+\frac{k_c}{2}\big)V^{2}/\hbar$ used in
\eqref{eq:K2-corrected}.

\medskip

Solving \eqref{eq:CS-N-1D-c} by Fourier transform in $t$ and characteristics,
one finds that the integration constants are functions of the relative
coordinates $\mathbf r$, constant only with respect to $U$. The resulting
kernel for $U\neq 0$ is
\begin{equation}
\begin{aligned}
K_{N}(U,\mathbf r;\,t,t')
&=\sqrt{\frac{m c^{3} N}{2\pi i\hbar\,U}}\,
\exp\!\Bigg[
-\frac{i m c^{3} N}{2\hbar U}
\Big((t{-}t_{0})-(t'{-}t_{0})
-\frac{k_{N}\,U^{3}}{6 m c^{3} N^{2}}
-\frac{C(\mathbf r)\,U^{2}}{3c}\Big)^{2} \\
&\hspace{7.5em}
-\frac{i\,k_{N}^{2}U^{5}}{40\,\hbar\,m c^{3}\,N^{3}}
+\frac{i\,k_{N}}{2\hbar N}\,U^{2}(t{-}t_{0})
\Bigg],
\end{aligned}
\label{eq:KN}
\end{equation}
which reduces to \eqref{eq:K2-corrected} when $N=2$, $U=x_1+x_2$ and
$\mathbf r=(V)$.

Prescribing boundary data on the hyperplane $U=0$,
\[
\Phi_{0}(\mathbf r,t):=\Phi(U=0,\mathbf r,t),
\]
the \emph{general solution} can be written as
\begin{equation}
\Phi(\mathbf x,t)
=\int_{-\infty}^{\infty}\!dt'\;
K_{N}(U,\mathbf r;\,t,t')\,\Phi_{0}(\mathbf r,t'),
\qquad U=\sum_{j=1}^{N}x_j.
\label{eq:Phi-N-kernel}
\end{equation}

To explore the dynamics in closed form, we choose Gaussian temporal data at
$U=0$ with an arbitrary relative profile $f(\mathbf r)$,
\begin{equation}
\Phi_{0}(\mathbf r,t)
= f(\mathbf r)\,
  \frac{1}{(2\pi\sigma^2)^{1/4}}
  \exp\!\left[-\frac{(t-t_0)^{2}}{4\sigma^{2}}\right].
\label{eq:Phi0-N-Gauss}
\end{equation}
Substituting \eqref{eq:Phi0-N-Gauss} into \eqref{eq:Phi-N-kernel}, the integral
remains Gaussian and the field is
\begin{equation}
\begin{aligned}
\Phi(\mathbf x,t)
&=f(\bm r)\,\frac{1}{(2\pi)^{1/4}\sqrt{\Sigma_{N}(U)}}
\exp\!\left[-\frac{\big((t{-}t_{0})-t_{c,N}(U,\mathbf r)\big)^{2}}
{4\Sigma_{N}^{2}(U)}\right] \\
&\quad\times
\exp\!\left\{i\Big[\chi_{N}(U)\big((t{-}t_{0})-t_{c,N}(U,\mathbf r)\big)^{2}
+\frac{k_{N}}{2\hbar N}\,U^{2}(t{-}t_{0})
+\varphi_{N}(U,\mathbf r)\Big]\right\},
\end{aligned}
\label{eq:Phi-N-Gauss}
\end{equation}
where
\begin{equation}
\Sigma_{N}^{2}(U)=\sigma^{2}+\frac{\hbar^{2}U^{2}}{4\sigma^{2}m^2 c^6 N^{2}},\quad
t_{c,N}(U,\mathbf r)
=\frac{k_{N}\,U^{3}}{6 m c^{3} N^{2}}
+\frac{C(\mathbf r)\,U^{2}}{3c},\quad
\chi_{N}(U)=\frac{\hbar\,U}{8\sigma^{2}m c^3\,N\,\Sigma_{N}^{2}(U)}.
\label{eq:Sigma-tc-chi-N}
\end{equation}
The function $\varphi_{N}(U,\mathbf r)$ is a real, $t$–independent phase
collecting the remaining kernel and normalization contributions; its explicit
form is not needed for the densities and currents, and for $N=2$ it reduces to
the phase $\varphi(U,V)$ in \eqref{eq:Phi-2body-Gauss-corrected}.

\medskip

Returning to $\Psi$ via the gauge
\begin{equation}
\Psi(\mathbf x,t)=\exp\!\Big(\tfrac{i}{\hbar}(t-t_0)\,U_{\!tot}(\mathbf x)\Big)\,
\Phi(\mathbf x,t),
\qquad |\Psi|^{2}=|\Phi|^{2},
\label{eq:Psi-from-Phi-N}
\end{equation}
we obtain the $N$–body temporal density
\begin{equation}
\rho_{t}^{(N)}(\bm x,t)=|\Psi|^{2}
=\frac{|f(\bm r)|^{2}}{\sqrt{2\pi}\,\Sigma_{N}(U)}\,
\exp\!\left[-\frac{\big((t{-}t_{0})-t_{c,N}(U,\mathbf r)\big)^{2}}
{2\Sigma_{N}^{2}(U)}\right],
\label{eq:rhoS-N}
\end{equation}
and the corresponding gauge–invariant temporal current
\begin{equation}
J^{(N)}_{t}(\bm x,t)
=\rho_{t}^{(N)}(\bm x,t)\left[
\frac{\omega^{2}U^{2}}{N c^{3}}
+\frac{C(\mathbf r)}{c^{2}}
+\frac{\hbar^{2}}{2 m c^{4}\sigma^{2}N}\,
\frac{U}{\Sigma_{N}^{2}(U)}\,
\big((t{-}t_{0})-t_{c,N}(U,\mathbf r)\big)\right],
\qquad U=\sum_{j=1}^{N}x_j .
\label{eq:J-N-closed}
\end{equation}
For $N=2$ and $\mathbf r=(V)$, the expressions
\eqref{eq:Phi-N-Gauss}–\eqref{eq:J-N-closed} reduce to the two–body formulas
\eqref{eq:Phi-2body-Gauss-corrected},
\eqref{eq:rhoS-2-Psi-corrected} and
\eqref{eq:J-2-closed-corrected}, with $C(\mathbf r)\to C(V)$.

\subsection{Model case 2: Coulombic interactions}
\label{subsec:spatial-coulomb}

For $N$ particles on a line, consider the (regularized) Coulomb potential
\begin{equation}
  U_{C}(\mathbf x)\;=\;-\sum_{1\le j<k\le N}\frac{k_e\,q_j q_k}{|x_j-x_k|},
  \qquad
  \partial_{x_j}U_{C}(\mathbf x)
  \;=\;\sum_{k\ne j}\frac{k_e\,q_j q_k}{|x_j-x_k|^{3}}\,(x_j-x_k).
  \label{eq:U-coulomb-1D}
\end{equation}
We tacitly assume a short–distance regularization so that all derivatives are
bounded at $x_j=x_k$.

In the gauge–transformed outside–the–square equation
\eqref{eq:CS-outside-space-only}, the Coulomb potential enters only through the
combination $\sum_{j}\partial_{x_j}U_{\!tot}(\mathbf x)$. For the purely
Coulombic case $U_{\!tot}=U_{\mathrm C}$, this sum vanishes identically.
Indeed,
\begin{equation}
  \sum_{j=1}^{N}\partial_{x_j}U_{\mathrm C}(\mathbf x)
  = - \sum_{1\le j<k\le N}k_e q_j q_k
    \Big(\partial_{x_j}\frac{1}{|x_j-x_k|} + \partial_{x_k}\frac{1}{|x_j-x_k|}\Big).
\end{equation}

each pair $(j,k)$ contributes zero in the sum,
\begin{equation}
  \partial_{x_j}\frac{1}{|x_j-x_k|}
  +\partial_{x_k}\frac{1}{|x_j-x_k|}=0
  \quad\Longrightarrow\quad
  \sum_{j=1}^{N}\partial_{x_j}U_{\mathrm C}(\mathbf x)=0.
\end{equation}
This is Newton’s third law in disguise: internal Coulomb forces add up to
zero in the center-of-mass direction. As a result, for purely Coulombic
interactions the gauge–transformed Carroll--Schr\"odinger equation
\eqref{eq:CS-outside-space-only} reduces to the \emph{free} Carroll-Schrödinger equation
\begin{equation}
  i\hbar c\sum_{j=1}^{N}\partial_{x_j}\Phi(\mathbf x,t)
  \;=\;-\frac{\hbar^{2}}{2mc^{2}}\;\partial_t^{2}\Phi(\mathbf x,t),
  \label{eq:CS-free-Coulomb}
\end{equation}
while the original field is obtained by the $t$–dependent gauge
\begin{equation}
  \Psi(\mathbf x,t)
  =\exp\!\Big(\tfrac{i}{\hbar}(t-t_0)\,U_{\mathrm C}(\mathbf x)\Big)\,\Phi(\mathbf x,t),
  \qquad
  |\Psi|^{2}=|\Phi|^{2}.
  \label{eq:Psi-from-Phi-Coulomb}
\end{equation}
Thus, Coulomb interactions enter the collective evolution through the phase and
through the choice of boundary data, while the bulk kernel governing the
center-of-mass direction coincides with that of a free system.

\emph{This model highlights an important limitation of multi-particle Carroll--Schr\"odinger systems with purely spatial, translation-invariant internal interactions. Whenever $U_{\!tot}(\mathbf x)$ is invariant under global translations, the internal forces cancel in the collective drive $\sum_j\partial_{x_j}U_{\!tot}(\mathbf x)$ and do not generate additional collective driving terms beyond those already present in the free case. As a result, once the boundary/initial data on the collective hyperplane are fixed, the subsequent collective evolution is governed by the same kernel as in the free system, and the detailed relative-coordinate dynamics is not directly resolved by the collective CS equation. Physically, this behavior is a direct manifestation of Carrollian ultralocality. In the limit $c \to 0$, light cones collapse and spatial points become causally disconnected, implying that spatial gradients are suppressed and interactions become ultralocal in space \cite{deBoer2022, deBoerStories}. Consequently, a purely spatial potential cannot mediate propagation between separated points in the way it does in non-relativistic mechanics; the ``freezing'' of the relative spatial configuration is thus not a defect of the model, but a feature of the underlying Carrollian geometry.}

\section{From the many-body time-independent Schr\"odinger equation
to a space-independent many-body Carroll--Schr\"odinger equation}
\label{sec:multi_Sch_to_Carroll}

In our previous one-body analysis \cite{Rojas2025} we showed that, for a
space-independent Carroll potential $V_{\mathrm{car}}(t)$, the
space-independent Carroll--Schr\"odinger equation can be related to a
time-independent Schr\"odinger problem by a reparametrization
$x=\delta(t)$, with the two potentials linked through a Schwarzian
relation. Here we extend this construction to a simple but useful
many-body setting, namely an $N$-particle time-independent Schr\"odinger
equation with a \emph{separable} potential. This suffices to build a
space-independent many-body Carroll system on
$L^{2}(\mathbb R^{N}_{\bm t})$, and provides a convenient bridge between
stationary Schr\"odinger models and the temporal Carroll framework
developed in Sec.~\ref{sec:time-exchange-HBT}.

Consider $N$ identical particles of mass $m$ on the line with
a separable static potential,
\begin{equation}
  \bigg[
    -\,\frac{\hbar^{2}}{2m}\sum_{i=1}^{N}\partial_{x_i}^{2}
    + \sum_{i=1}^{N}V_{\mathrm{sch}}^{(i)}(x_i)
  \bigg]\Psi_{\mathrm{sch}}(\bm x)
  = E_{\mathrm{sch}}\,\Psi_{\mathrm{sch}}(\bm x),
  \qquad
  \bm x=(x_1,\ldots,x_N),
  \label{eq:TI_Sch_multi}
\end{equation}
where $V_{\mathrm{sch}}^{(i)}$ are one-body potentials. (More general
interacting N-body non-separable potentials require genuinely multivariable
reparametrizations and will not be considered here.) In the separable
case, we may seek stationary product states of the form
\begin{equation}
  \Psi_{\mathrm{sch}}(\bm x)
  =
  \prod_{i=1}^{N}\psi_{i}(x_i),
  \qquad
  E_{\mathrm{sch}}=\sum_{i=1}^{N}E_{\mathrm{sch}}^{(i)},
  \label{eq:TI_Sch_multi_prod}
\end{equation}
with each factor solving a one-dimensional time-independent
Schr\"odinger equation
\begin{equation}
  -\,\frac{\hbar^{2}}{2m}\,\psi_{i}''(x_i)
  + V_{\mathrm{sch}}^{(i)}(x_i)\,\psi_{i}(x_i)
  = E_{\mathrm{sch}}^{(i)}\,\psi_{i}(x_i),
  \qquad i=1,\ldots,N.
  \label{eq:TI_Sch_1D_i}
\end{equation}
For indistinguishable particles, bosonic or fermionic symmetry can be
enforced by symmetrizing or antisymmetrizing the products
\eqref{eq:TI_Sch_multi_prod}; the mapping constructed below acts
coordinate-wise and may be extended to these symmetrized sectors.

On the Carroll side we consider space-independent many-body generators
built from $N$ temporal configuration variables
$\bm t=(t_1,\ldots,t_N)$ on equal-$x$ slices, with one-body Carroll
potentials $V_{\mathrm{car}}^{(i)}(t_i)$,
\begin{equation}
  \bigg[
    \sum_{i=1}^{N}\frac{1}{2mc^{2}}
    \Big(-i\hbar\,\partial_{t_i}-V_{\mathrm{car}}^{(i)}(t_i)\Big)^{2}
    - cP_{0}
  \bigg]\Phi(\bm t) = 0,
  \label{eq:Car_space_indep_multi}
\end{equation}
where $P_{0}$ is a constant (the total Carroll momentum along the
evolution coordinate $x$). Just as in the one-body case, it is useful to
parametrize $P_{0}$ in terms of positive ``energy labels'' $E_{0}^{(i)}$,
\begin{equation}
  c p_{0}^{(i)}=\frac{[E_{0}^{(i)}]^{2}}{2mc^{2}},
  \qquad
  P_{0}=\sum_{i=1}^{N}p_{0}^{(i)},
  \label{eq:p0_multi_def}
\end{equation}
so that the multi-time equation \eqref{eq:Car_space_indep_multi} admits a
separated solution of the form
\begin{equation}
  \Phi(\bm t)
  = \prod_{i=1}^{N}\phi_{i}(t_i),
  \qquad
  \frac{1}{2mc^{2}}\Big(-i\hbar\,\frac{d}{dt_i}-V_{\mathrm{car}}^{(i)}(t_i)\Big)^{2}\phi_{i}(t_i)
  - c p_{0}^{(i)}\,\phi_{i}(t_i)=0.
  \label{eq:Car_space_indep_1D_i}
\end{equation}
As in the one-body discussion, the defining separation parameters are
the Carroll momenta $p_{0}^{(i)}$; the corresponding labels
$E_{0}^{(i)}=\pm\sqrt{2mc^{3}p_{0}^{(i)}}$ are two-to-one functions of
$p_{0}^{(i)}$.

The goal of this section is to exhibit, for each $i$, a coordinate
reparametrization $x_i=\delta_{i}(t_i)$ relating
\eqref{eq:TI_Sch_1D_i} and \eqref{eq:Car_space_indep_1D_i}, and then to
assemble these one-dimensional maps into a many-body correspondence
between \eqref{eq:TI_Sch_multi} and \eqref{eq:Car_space_indep_multi}.

\subsection{Coordinate reparametrization and one-body mapping}

We now recall the one-body construction of \cite{Rojas2025}, adapted to
the $i$th coordinate. Let
\begin{equation}
  x_i=\delta_{i}(t_i),\qquad
  \psi_{i}(x_i)\;\longrightarrow\;
  \phi_{i}(t_i):=\psi_{i}\!\big(\delta_{i}(t_i)\big),
  \label{eq:map_delta_i_def}
\end{equation}
with $\delta_{i}$ a smooth, monotone reparametrization. Writing
$\dot\delta_{i}=d\delta_{i}/dt_i$, a direct application of the chain
rule gives
\begin{equation}
  \partial_{x_i}\psi_{i}
  = \frac{1}{\dot\delta_{i}}\partial_{t_i}\phi_{i},
  \qquad
  \partial_{x_i}^{2}\psi_{i}
  = \frac{1}{\dot\delta_{i}^{2}}\partial_{t_i}^{2}\phi_{i}
    - \frac{\ddot\delta_{i}}{\dot\delta_{i}^{3}}\partial_{t_i}\phi_{i}.
\end{equation}
Substituting into the time-independent Schr\"odinger equation
\eqref{eq:TI_Sch_1D_i} yields, after a simple rearrangement,
\begin{equation}
  -\frac{\hbar^{2}}{2m}
  \left[
    \frac{\ddot\phi_{i}}{\dot\delta_{i}^{2}}
    - \frac{\dot\phi_{i}\ddot\delta_{i}}{\dot\delta_{i}^{3}}
  \right]
  + V_{\mathrm{sch}}^{(i)}\!\big(\delta_{i}(t_i)\big)\,\phi_{i}(t_i)
  - E_{\mathrm{sch}}^{(i)}\,\phi_{i}(t_i)
  = 0.
  \label{eq:Sch_transformed_i}
\end{equation}
Multiplying \eqref{eq:Sch_transformed_i} by
$\dot\delta_{i}^{2}/c^{2}$ and comparing with the expanded
space-independent Carroll equation
\eqref{eq:Car_space_indep_1D_i},
\begin{equation}
  -\frac{\hbar^{2}}{2mc^{2}}\ddot\phi_{i}
  + \frac{i\hbar\,V_{\mathrm{car}}^{(i)}(t_i)}{mc^{2}}\dot\phi_{i}
  + \Bigg[
      \frac{i\hbar}{2mc^{2}}\frac{dV_{\mathrm{car}}^{(i)}}{dt_i}
      + \frac{\big(V_{\mathrm{car}}^{(i)}(t_i)\big)^{2}}{2mc^{2}}
      - c p_{0}^{(i)}
    \Bigg]\phi_{i}(t_i)
  =0,
  \label{eq:Car_expand_i}
\end{equation}
we obtain the one-dimensional relations
\begin{equation}
  \frac{\ddot\delta_{i}}{\dot\delta_{i}}
  = \frac{2i}{\hbar}\,V_{\mathrm{car}}^{(i)}(t_i),
  \qquad
  V_{\mathrm{sch}}^{(i)}\!\big(\delta_{i}(t_i)\big)
  = E_{\mathrm{sch}}^{(i)}
    + \frac{1}{2m\,\dot\delta_{i}^{2}}
      \left[
        i\hbar\,\frac{dV_{\mathrm{car}}^{(i)}}{dt_i}
        + \big(V_{\mathrm{car}}^{(i)}(t_i)\big)^{2}
        - \big(E_{0}^{(i)}\big)^{2}
      \right].
  \label{eq:map_start_multi}
\end{equation}
The first equation integrates to
\begin{equation}
  \delta_{i}(t_i)
  = C_{1}^{(i)}
    + C_{0}^{(i)}\int^{t_i}\!
       \exp\!\left(\frac{2i}{\hbar}
       \int^{t'}\!V_{\mathrm{car}}^{(i)}(s)\,ds\right)dt',
  \label{eq:delta_integrated_i}
\end{equation}
with any lower limits absorbed into the constants
$C_{0}^{(i)},C_{1}^{(i)}$. Eliminating $V_{\mathrm{car}}^{(i)}$ from
\eqref{eq:map_start_multi} via
$V_{\mathrm{car}}^{(i)}=-\tfrac{i\hbar}{2}\,\ddot\delta_{i}/\dot\delta_{i}$ leads to
the Schwarzian form
\begin{equation}
  V_{\mathrm{sch}}^{(i)}\!\big(\delta_{i}(t_i)\big)
  - E_{\mathrm{sch}}^{(i)}
  =
  \frac{\hbar^{2}}{4m}
  \frac{\{\delta_{i},t_i\}}{\dot\delta_{i}^{2}}
  - \frac{\big(E_{0}^{(i)}\big)^{2}}{2m}\,\frac{1}{\dot\delta_{i}^{2}},
  \qquad
  \{\delta_{i},t_i\}
  :=\frac{\delta_{i}'''}{\delta_{i}'}
    -\frac{3}{2}\!\left(\frac{\delta_{i}''}{\delta_{i}'}\right)^{2}.
  \label{eq:forward_S_multi_i}
\end{equation}
Writing $\tau_{i}(x_i):=\delta_{i}^{-1}(x_i)$ and using the inversion
identities
\(
(\{\delta_{i},t_i\}/\dot\delta_{i}^{2})|_{t_i=\tau_{i}(x_i)}=-\{\tau_{i},x_i\},
\;
(1/\dot\delta_{i}^{2})|_{t_i=\tau_{i}(x_i)}=\tau_{i}'(x_i)^{2},
\)
we arrive at the inverse equation
\begin{equation}
  \boxed{\;
  \{\tau_{i},x_i\}
  +\frac{2\big(E_{0}^{(i)}\big)^{2}}{\hbar^{2}}\,\tau_{i}'(x_i)^{2}
  = -\,\frac{4m}{\hbar^{2}}
    \Big(V_{\mathrm{sch}}^{(i)}(x_i)-E_{\mathrm{sch}}^{(i)}\Big),
  \qquad i=1,\ldots,N.
  \;}
  \label{eq:inverse_master_multi_i}
\end{equation}

Equation \eqref{eq:inverse_master_multi_i} is the many-body analogue,
coordinate by coordinate, of the one-body inverse relation
\cite{Rojas2025}. As before, it can be linearized by
using the Schwarzian chain rule. For each $i$ we pick a function
$f_{i}$ with constant Schwarzian
$\{f_{i},u\}=2(E_{0}^{(i)}/\hbar)^{2}$, for instance
$f_{i}(u)=\tan\!\big(\tfrac{E_{0}^{(i)}}{\hbar}u\big)$, and define
\begin{equation}
  \sigma_{i}(x_i):=f_{i}\!\big(\tau_{i}(x_i)\big)
  =\tan\!\Big(\tfrac{E_{0}^{(i)}}{\hbar}\,\tau_{i}(x_i)\Big),
\end{equation}
so that $\{\sigma_{i},x_i\}=\{\tau_{i},x_i\}+\tfrac{2(E_{0}^{(i)})^{2}}{\hbar^{2}}\tau_{i}'^{2}$. Then
\eqref{eq:inverse_master_multi_i} reduces to the pure Schwarzian
equation
\begin{equation}
  \boxed{\;
  \{\sigma_{i},x_i\}
  =-\,\frac{4m}{\hbar^{2}}
    \Big(V_{\mathrm{sch}}^{(i)}(x_i)-E_{\mathrm{sch}}^{(i)}\Big),
  \qquad i=1,\ldots,N.
  \;}
  \label{eq:pure_S_multi_i}
\end{equation}
A standard construction shows that if $y_{1}^{(i)},y_{2}^{(i)}$ form a
fundamental system of the linear ODE
\begin{equation}
  y''(x_i)+q_{i}(x_i)\,y(x_i)=0,
  \qquad
  q_{i}(x_i)
  :=\frac{2m}{\hbar^{2}}
  \Big(V_{\mathrm{sch}}^{(i)}(x_i)-E_{\mathrm{sch}}^{(i)}\Big),
  \label{eq:lin_ode_q_multi_i}
\end{equation}
then $\sigma_{i}(x_i)=y_{1}^{(i)}(x_i)/y_{2}^{(i)}(x_i)$ satisfies
$\{\sigma_{i},x_i\}=-2q_{i}(x_i)$ and hence \eqref{eq:pure_S_multi_i}.
Thus, on any interval where $y_{2}^{(i)}$ has no zeros and $\tau_{i}$ is
monotone,
\begin{equation}
  \sigma_{i}(x_i)
  =\frac{y_{1}^{(i)}(x_i)}{y_{2}^{(i)}(x_i)},
  \qquad
  \tau_{i}(x_i)
  =\frac{\hbar}{E_{0}^{(i)}}\arctan\!\big(\sigma_{i}(x_i)\big),
  \qquad
  \boxed{\,\delta_{i}(t_i)=\tau_{i}^{-1}(t_i)\,}.
  \label{eq:delta_construct_multi_i}
\end{equation}
Substituting \eqref{eq:delta_construct_multi_i} into
\eqref{eq:forward_S_multi_i} and \eqref{eq:map_start_multi} reproduces
the original $V_{\mathrm{sch}}^{(i)}(x_i)$ and determines the associated
Carroll potential $V_{\mathrm{car}}^{(i)}(t_i)$ via
$V_{\mathrm{car}}^{(i)}=-\tfrac{i\hbar}{2}\ddot\delta_{i}/\dot\delta_{i}$.

\subsection{A preliminary interacting example: two coupled harmonic oscillators}

As an illustration of how an interacting static potential can be incorporated into this framework, we consider the two-body coupled harmonic oscillator potential
\begin{equation}
  V_{\mathrm{sch}}^{(2)}(x_1,x_2)
  = \frac{1}{2}m\omega^{2}(x_1^{2}+x_2^{2})
    + \frac{1}{2}k_{c}(x_1-x_2)^{2}.
  \label{eq:Vsch-2osc}
\end{equation}
By introducing center-of-mass and relative coordinates, $X = (x_1+x_2)/2$ and $\xi = x_1-x_2$, the kinetic energy separates into terms with total mass $M=2m$ and reduced mass $\mu=m/2$. The potential energy diagonalizes similarly,
\begin{equation}
  V_{\mathrm{sch}}^{(2)}(X,\xi)
  = \frac{1}{2}M\Omega_{X}^{2}X^{2}
    + \frac{1}{2}\mu\Omega_{\xi}^{2}\xi^{2},
\end{equation}
with effective frequencies given by $\Omega_{X} = \omega$ and $\Omega_{\xi} = \sqrt{\omega^{2}+2k_c/m}$.

The stationary problem thus factorizes into two independent one-dimensional channels. We can therefore apply the coordinate duality defined by Eq.~\eqref{eq:pure_S_multi_i} independently to each mode. This yields two reparametrization functions, $X=\delta_{X}(t_{X})$ and $\xi=\delta_{\xi}(t_{\xi})$, determined by the Schwarzian equations for the effective potentials $V_{\mathrm{sch}}^{(X)} = \frac{1}{2}M\Omega_X^2 X^2$ and $V_{\mathrm{sch}}^{(\xi)} = \frac{1}{2}\mu\Omega_\xi^2 \xi^2$.

To recover the relationship between the physical coordinates and the particle times, we invert the spatial linear transformation and substitute the temporal normal modes $t_{X}=(t_1+t_2)/2$ and $t_{\xi}=t_1-t_2$. This yields the mixed map
\begin{align}
  x_1 &= \delta_{X}\!\left(\frac{t_1+t_2}{2}\right) + \frac{1}{2}\delta_{\xi}(t_1-t_2),
  \label{eq:map-x1-coupled}
  \\[0.4em]
  x_2 &= \delta_{X}\!\left(\frac{t_1+t_2}{2}\right) - \frac{1}{2}\delta_{\xi}(t_1-t_2).
  \label{eq:map-x2-coupled}
\end{align}
Finally, the resulting Carroll generator is defined by the sum of the effective potentials evaluated at these temporal arguments
\begin{equation}
  V_{\mathrm{car}}^{(2)}(t_1,t_2)
  =
  V_{\mathrm{car}}^{(X)}\!\Big(\frac{t_1+t_2}{2}\Big)
  + V_{\mathrm{car}}^{(\xi)}(t_1-t_2).
  \label{eq:Vcar-2osc-mixed}
\end{equation}
This example demonstrates that spatial entanglement arising from potential coupling in the Schr\"odinger picture maps to a mixing of temporal coordinates in the Carrollian picture, mediated by the mode-dependent reparametrizations.

\section{Exchange symmetry in the time domain and Hanbury Brown--Twiss correlations}
\label{sec:time-exchange-HBT}

\subsection{Exchange symmetry of multiparticle CS equation}

On each equal-$x$ slice the $N$-body state is
\[
\Psi(x;\bm t)\in L^2(\mathbb R_{\bm t}^N),
\qquad
\bm t=(t_1,\ldots,t_N),
\]
as in Ref.~\cite{Rojas2025}. Since the Carroll--Schr\"odinger (CS) equation
evolves in $x$ and treats $t$ as the configuration coordinate, exchange
acts on the time labels $t_1,\dots,t_N$ within each equal-$x$ slice,
just as it acts on spatial labels on equal-time slices in standard
Schr\"odinger theory with stationary fields. For indistinguishable
particles, physical probabilities must be invariant under any
permutation $\pi\in S_N$,
\[
\|\Psi(x;\bm t)\|^2
=\int d\bm t\,|\Psi(x;\bm t)|^2
=\int d\bm t\,|\Psi(x;t_{\pi(1)},\ldots,t_{\pi(N)})|^2,
\]
so there is a unitary representation $U_\pi$ acting as
\[
(U_\pi\Psi)(x;\bm t)=\Psi(x;t_{\pi(1)},\ldots,t_{\pi(N)}).
\]
If the statistics sector is one-dimensional under $U_\pi$, then
$U_\pi\Psi=\chi(\pi)\Psi$ with $|\chi(\pi)|=1$. Since transpositions
$\tau_{ij}$ generate $S_N$ and $\tau_{ij}^2=1$, one has
\[
\chi(\tau_{ij})^2=1\quad\Longrightarrow\quad \chi(\tau_{ij})\in\{+1,-1\},
\]
which extends to a character of all $S_N$. This leaves the two sectors
\[
\Psi_{\mathrm B}(x;\bm t)
=+\Psi_{\mathrm B}(x;t_{\pi(1)},\ldots,t_{\pi(N)}),\qquad
\Psi_{\mathrm F}(x;\bm t)
=-\Psi_{\mathrm F}(x;t_{\pi(1)},\ldots,t_{\pi(N)}),
\quad\forall\pi\in S_N.
\]
These are the bosonic and fermionic sectors \emph{in the time domain}:
``fermionic in time’’ means antisymmetry under $t_i\leftrightarrow t_j$
at fixed $x$, the direct analogue of exchange acting on spatial labels
at fixed time.

For later use we introduce the shorthand
\[
\bm t_{a:b}:=(t_a,t_{a+1},\ldots,t_b),
\qquad
d\bm t_{a:b}:=dt_a\,dt_{a+1}\cdots dt_b.
\]
With this notation, the equal-$x$ pair density
\begin{equation}
n^{(2)}(t,t')
=
N(N{-}1)\!\int_{\mathbb R^{N-2}}
\big|\Psi(x;\,t,t',\bm t_{3:N})\big|^2\,d\bm t_{3:N}
\label{eq:pair-density-def}
\end{equation}
is the reduced density for finding one time at $t$ and another at
$t'$ on the slice. For a fermionic-in-time state, antisymmetry under
exchanging any two time labels enforces nodes on the coincidence
hyperplanes $t=t'$, i.e.
\[
\Psi(x;\ldots,t,\ldots,t,\ldots)=0,
\]
and therefore $n^{(2)}(t,t)=0$. This is Pauli suppression
\emph{in time} (temporal antibunching).

To quantify temporal coherence on a slice we introduce the equal-$x$
one-body reduced density matrix (1RDM),
\begin{equation}
\gamma(t,t')
=
N\!\int_{\mathbb R^{N-1}}
\Psi(x;\,t,\bm t_{2:N})\,
\Psi(x;\,t',\bm t_{2:N})^{*}\,
d\bm t_{2:N},
\qquad
n(t)=\gamma(t,t),
\label{eq:1rdm-def}
\end{equation}
whose diagonal $n(t)$ is the one-body temporal density. The kernel
$\gamma$ is Hermitian and positive semidefinite, with
$\mathrm{Tr}\,\gamma=\int n(t)\,dt=N$.

\subsection{Hanbury Brown--Twiss correlations}

The pair density and 1RDM also encode an equal-$x$ analogue of
Hanbury Brown--Twiss (HBT) correlations \cite{HBT1956,MandelWolf1995}. On a given slice we interpret
$n(t)\,dt$ as the single-particle detection density in the interval
$(t,t+dt)$, while $n^{(2)}(t,t')\,dt\,dt'$ is the joint detection
density in $(t,t+dt)\times(t',t'+dt')$. This motivates the normalized
temporal second-order coherence function
\begin{equation}
  g^{(2)}(t,t')
  :=\frac{n^{(2)}(t,t')}{n(t)\,n(t')}.
  \label{eq:g2-def}
\end{equation}

It is convenient to express $g^{(2)}$ in terms of the 1RDM. Defining
the normalized first-order coherence
\begin{equation}
  g^{(1)}(t,t')
  :=\frac{\gamma(t,t')}{\sqrt{n(t)\,n(t')}},
  \label{eq:g1-def}
\end{equation}
one finds, for the noninteracting Bose and Fermi reference states
built from orthonormal time-orbitals, the standard Wick factorization
relations
\begin{align}
  n^{(2)}_{\mathrm F}(t,t')
  &= n(t)\,n(t') - |\gamma(t,t')|^{2},
  \label{eq:n2F-gamma}
  \\[0.4em]
  n^{(2)}_{\mathrm B}(t,t')
  &= n(t)\,n(t') + |\gamma(t,t')|^{2},
  \label{eq:n2B-gamma}
\end{align}
and therefore
\begin{align}
  g^{(2)}_{\mathrm F}(t,t')
  &= 1 - \frac{|\gamma(t,t')|^{2}}{n(t)\,n(t')}
   = 1 - |g^{(1)}(t,t')|^{2},
  \label{eq:g2F-g1}
  \\[0.4em]
  g^{(2)}_{\mathrm B}(t,t')
  &= 1 + \frac{|\gamma(t,t')|^{2}}{n(t)\,n(t')}
   = 1 + |g^{(1)}(t,t')|^{2}.
  \label{eq:g2B-g1}
\end{align}
On the coincidence line $t'=t$, using $\gamma(t,t)=n(t)$, we obtain
\begin{equation}
  g^{(2)}_{\mathrm F}(t,t)=0,
  \qquad
  g^{(2)}_{\mathrm B}(t,t)
  =1+\frac{|\gamma(t,t)|^{2}}{n(t)^{2}}
  =2,
  \label{eq:g2-diagonal}
\end{equation}
in agreement with the Pauli suppression in time inferred from
Eq.~\eqref{eq:pair-density-def}. Ideal Carroll fermions in time
exhibit perfect temporal antibunching, while ideal Carroll bosons
display temporal bunching with $g^{(2)}(t,t)=2$ on each equal-$x$
slice. Away from the coincidence line, the modulus of $g^{(1)}$ in
Eqs.~\eqref{eq:g2F-g1}--\eqref{eq:g2B-g1} measures the temporal
coherence length of the Carroll gas.

These expressions suggest a Carrollian HBT protocol. One prepares the
system in a fixed equal-$x$ state $\Psi(x_{0};\bm t)$, lets it
propagate according to CS dynamics, and places a detector at a
downstream position $x$. On each run the detector records a list of
arrival times $\{t_{1}^{(r)},\ldots,t_{M_{r}}^{(r)}\}$, from which
empirical estimates of $n(t)$, $n^{(2)}(t,t')$, and hence
$g^{(2)}(t,t')$ can be constructed. In the noninteracting limit the
outcomes are governed by
Eqs.~\eqref{eq:g2F-g1}--\eqref{eq:g2-diagonal}; temporal interactions
introduced later deform $\gamma$ and $n^{(2)}$, thus modifying the
shape of $g^{(2)}(t,t')$. Temporal HBT measurements at fixed $x$ then
provide an operational probe of many-body structure and statistics
\emph{in the time domain} for Carrollian quantum systems.

\section{From many-body CS to a temporal DNLS}

We work on equal-$x$ slices in second quantization. The field operators 
satisfy the canonical equal-$x$ relations \cite{Najafizadeh2025-2}
\begin{equation}
\big[\hat\psi(x,t),\hat\psi^\dagger(x,t')\big]=\frac{1}{c}\,\delta(t{-}t'),
\qquad
\big[\hat\psi(x,t),\hat\psi(x,t')\big]=\big[\hat\psi^\dagger(x,t),\hat\psi^\dagger(x,t')\big]=0,
\label{eq:ccr-equalx}
\end{equation}
so that the Hamiltonian $\hat H(x)$ generates $x$-evolution in the
Heisenberg picture via
\begin{equation}
i\hbar\,c\,\frac{d \hat{\mathcal O}}{d x}
\;=\; \big[\hat{\mathcal O},\hat H(x)\big]
\;+\;i\hbar\,c\,\big(\tfrac{\partial \mathcal O}{\partial x} \big)_{\!\text{explicit}},
\qquad\text{with}\qquad
\boxed{\;\hat H(x)=c\,\hat P(x)\;}
\label{eq:heis-energy}
\end{equation}
where $\hat P(x)$ is the $x$-momentum generator on the equal-$x$
slice.  For historical reasons we use the ``energy'' generator
$\hat H(x)=c\,\hat P(x)$, rather than $\hat P(x)$ itself, as the
basic generator of $x$-evolution \cite{Rojas2025}.
 Defining the Hamiltonian as
\begin{equation}
\label{eq:H-inside-field}
\begin{aligned}
\hat H(x)
=\;&\int c\,dt\;
:\hat\psi^\dagger(x,t)\,
\frac{1}{2mc^{2}}
\Big(\widehat E - \hat{\mathcal A}(x,t)\Big)^{2}
\hat\psi(x,t):\,,\\[0.4em]
\widehat E\;&=-i\hbar\,\partial_t, \qquad
\hat n(x,t)=\hat\psi^\dagger(x,t)\hat\psi(x,t),
\end{aligned}
\end{equation}
where colons indicate normal ordering and the temporal ``gauge field''
$\hat{\mathcal A}(x,t)$ encodes both one-body and two-body
interactions,
\begin{equation}
  \hat{\mathcal A}(x,t)
  := U(x,t)
   + c\!\int\!dt'\,w\!\big(x; t,t'\big)\,\hat n(x,t').
  \label{eq:A-hat-field}
\end{equation}
Here $U(x,t)$ is a real external temporal potential and
$w\!\big(x; t,t'\big)=w\!\big(x; t',t\big)$ is a real symmetric
two-time kernel representing short-range temporal interactions.  

The Heisenberg equation for the temporal field operator
follows from \eqref{eq:heis-energy} by choosing
$\hat{\mathcal O}=\hat\psi(x,t)$,
\begin{equation}
i\hbar c\,\partial_x \hat\psi(x,t)
\;=\;\big[\hat\psi(x,t),\hat H(x)\big].
\label{eq:heis-psi-inside}
\end{equation}
Because $\hat H(x)$ is quadratic in the shifted energy operator
$\widehat E-\hat{\mathcal A}$ and $\hat{\mathcal A}$ itself depends on
the density $\hat n$, evaluating the commutator requires repeated use
of the canonical relations \eqref{eq:ccr-equalx}.  In particular,
\begin{equation}
\big[\hat\psi(x,t),\hat n(x,t')\big]
=\frac{1}{c}\,\delta(t{-}t')\,\hat\psi(x,t),
\label{eq:psi-n-comm}
\end{equation}

To define a contact-in-time limit, we appeal to the physical constraints of the Carrollian regime. In the $c \to 0$ limit, finite propagation speeds are suppressed, and interactions tend toward ultralocality. We assume that the collision duration (or ``memory time'') of the interaction kernel is much shorter than the characteristic temporal coherence scale of the macroscopic envelope (this is the temporal analogue of the Lieb-Liniger potential \cite{LiebLiniger1963I}). Under this short-memory approximation, the kernel acts instantaneously
\begin{equation}
  w\!\big(x; t,t'\big)\ \longrightarrow\ g(x)\,\delta(t-t').
\end{equation}
This is the temporal analogue of the standard Fermi pseudopotential in spatial cold atom physics. In this limit, the temporal gauge field becomes local in $t$ at the operator level
\begin{equation}
  \hat{\mathcal A}(x,t)=U(x,t)+c\,g(x)\,\hat n(x,t),
  \label{eq:A-hat-local}
\end{equation}
with $g(x)$ representing the contact interaction strength. Substituting
\eqref{eq:A-hat-local} into \eqref{eq:H-inside-field} and using
\eqref{eq:ccr-equalx} and \eqref{eq:psi-n-comm}, the Heisenberg
equation \eqref{eq:heis-psi-inside} yields the operator evolution equation
\begin{equation}
\begin{aligned}
& i\hbar c\,\partial_x\hat\psi(x,t)
= -\frac{\hbar^2}{2mc^{2}}\partial_t^2\hat\psi(x,t)
  + \frac{i\hbar}{2mc^{2}}\big(\partial_t U(x,t)\big)\hat\psi(x,t)
 + \frac{i\hbar}{mc^{2}}\,U(x,t)\,\partial_t\hat\psi(x,t)
  +\\[0.4em]
& \frac{2i\hbar\,g(x)}{m c}\,\hat n(x,t)\,\partial_t\hat\psi(x,t)
  + \frac{1}{2mc^{2}}
    \Big[
      U(x,t)^{2}
      + 4c\,g(x)\,U(x,t)\,\hat n(x,t)
      + 3c^{2}g(x)^{2}\hat n(x,t)^{2}
    \Big]\hat\psi(x,t),
\end{aligned}
\label{eq:temporal-Heis-inside-local}
\end{equation}

Now taking the mean-field limit of the operator evolution equation
\begin{equation}
  \hat\psi(x,t)\;\to\;\phi(x,t),\qquad
  \hat n(x,t)\;\to\;|\phi(x,t)|^{2},\qquad
  \hat n(x,t)^{2}\;\to\;|\phi(x,t)|^{4},
\end{equation}
we arrive at the nonlinear evolution equation
\begin{equation}
\begin{aligned}
& i\hbar c\,\partial_x\phi(x,t)
= -\frac{\hbar^2}{2mc^{2}}\partial_t^2\phi(x,t)
  + \frac{i\hbar}{2mc^{2}}\big(\partial_t U(x,t)\big)\phi(x,t)
 + \frac{i\hbar}{mc^{2}}\,U(x,t)\,\partial_t\phi(x,t)
  +\\[0.4em]
& \frac{2i\hbar\,g(x)}{m c}\,|\phi(x,t)|^{2}\,\partial_t\phi(x,t)
  + \frac{1}{2mc^{2}}
    \Big[
      U(x,t)^{2}
      + 4c\,g(x)\,U(x,t)\,|\phi(x,t)|^{2}
      + 3c^{2}g(x)^{2}|\phi(x,t)|^{4}
    \Big]\phi(x,t),
\end{aligned}
\label{eq:temporal-GP-inside-full}
\end{equation}

In the simplest case of a purely interaction-driven temporal gas,
$U(x,t)=0$, the mean-field equation
\eqref{eq:temporal-GP-inside-full} reduces to
\begin{equation}
i\hbar c\,\partial_x\phi(x,t)
=-\frac{\hbar^2}{2mc^{2}}\partial_t^2\phi(x,t)
+\frac{2i\hbar\,g(x)}{m c}\,|\phi(x,t)|^{2}\,\partial_t\phi(x,t)
+\frac{3\,g(x)^{2}}{2m}\,|\phi(x,t)|^{4}\phi(x,t),
\label{eq:temporal-GP-inside-pure}
\end{equation}
with $|\phi|^{2}=|\phi(x,t)|^{2}$. For constant interaction strength $g(x)=g_{0}$,
this equation is structurally similar to the family of cubic--quintic derivative
nonlinear Schr\"odinger (DNLS) equations appearing in nonlinear optics and
plasma physics \cite{Mio1976,AgrawalNFO2019,JenkinsEtAl2020}.

To determine the specific coefficients, we bring the physical evolution equation
\eqref{eq:temporal-GP-inside-pure} into a dimensionless form using the rescaling
$x=L X$, $t=\tau T$, and $\phi=\mathcal{A}\psi$, where $\tau$ is the characteristic
pulse width. The dynamical scales are fixed by balancing the kinetic and nonlinear terms,
yielding the characteristic length $L = 2mc^3\tau^2/\hbar$ and field amplitude
$\mathcal{A} = \sqrt{\hbar/(4g_0 c \tau)}$. Substituting these scales reveals that
the quintic coefficient $\beta$ is strictly determined by the minimal coupling
structure of the theory. The resulting dimensionless equation is:
\begin{equation}
i\partial_{X}\psi + \partial_{T}^{2}\psi
- i|\psi|^{2}\partial_{T}\psi - \frac{3}{16} |\psi|^{4}\psi = 0.
\label{eq:derived_DNLS}
\end{equation}
We note that the negative imaginary coefficient of the cubic term arises directly
from the rescaling, though it can be mapped to the standard $+i$ form by a time
reflection $T \to -T$.

\begin{figure}[h]
    \centering
    \includegraphics[width=1.05\textwidth]{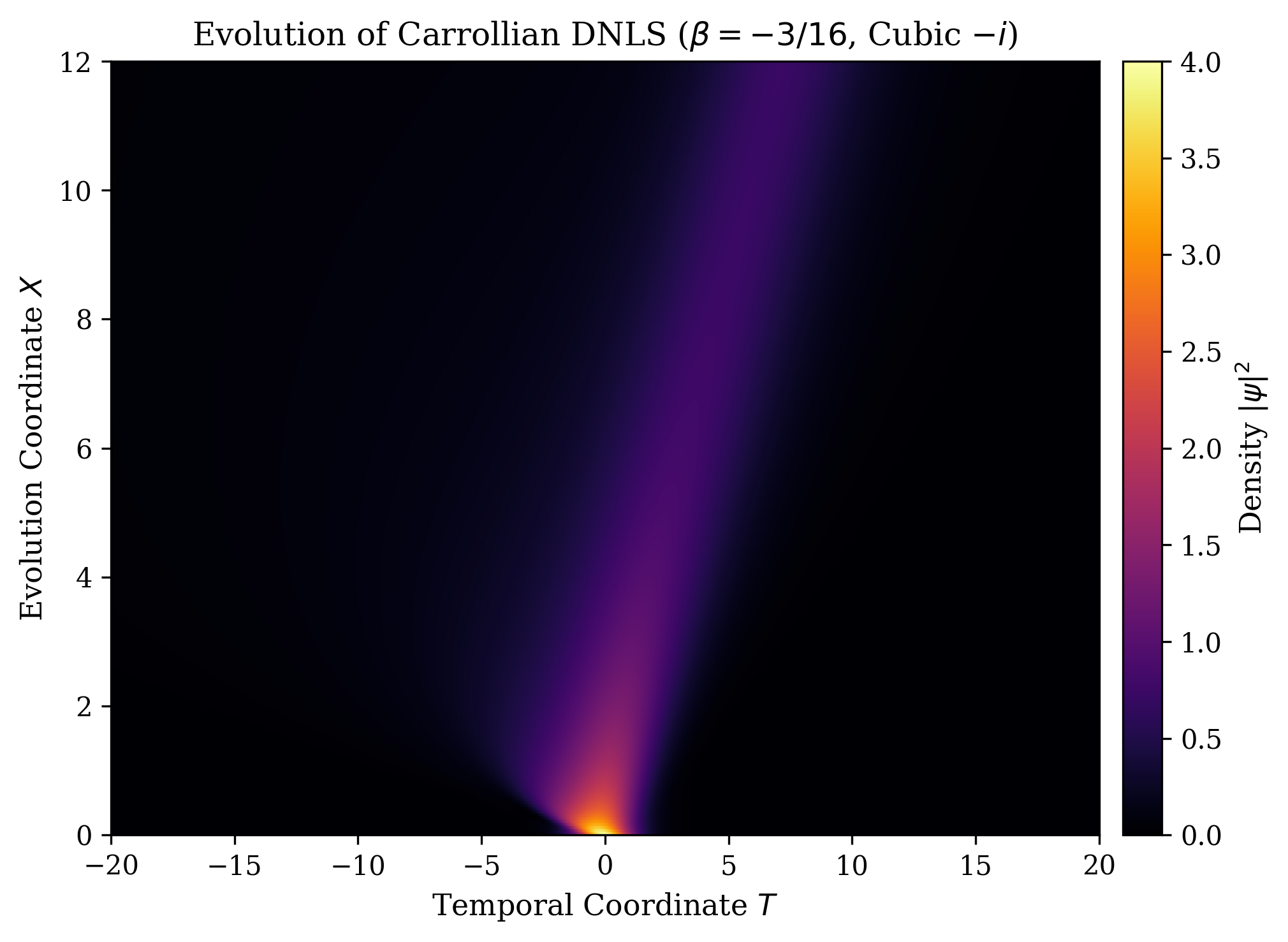}
    \caption{Numerical evolution of the interaction-driven Carroll gas density $|\psi(X,T)|^2$ for the derived parameters (cubic $-i$, quintic $\beta=-3/16$). The simulation demonstrates that the pulse remains strictly localized as it evolves along $X$. The dynamics are governed by the interplay between the derivative nonlinearity, which sustains the solitary wave structure, and the repulsive quintic term, which saturates the amplitude and ensures stability against collapse.}
    \label{fig:dnls_sim}
\end{figure}

The value $\beta = -3/16$ places the interaction-driven Carroll system in a specific
stability regime. While the minimal coupling generates a repulsive (defocusing) quintic term,
numerical analysis (Fig.~\ref{fig:dnls_sim}) reveals that this does not lead to
incoherent dispersion. Instead, the derivative cubic nonlinearity (which promotes
soliton formation) is sufficient to maintain the wave packet's coherence against dispersion. The result is the formation
of robust, localized solitary waves that propagate with a preserved profile over long
evolution distances.

\clearpage

\section{Carrollian Hohenberg--Kohn mapping and Kohn--Sham scheme}
\label{sec:Carroll-HK-KS}

We motivate the density--current functional framework by minimally coupling the canonical pairs $(x,P)$ and $(t_i,E_i)$ to external Abelian gauge fields on equal-$x$ slices. The momentum conjugate to $x$ couples to a scalar field $\Phi(t)$ via
\begin{equation}
  P \;\longrightarrow\; P - \frac{1}{c}\,\Phi(t),
\end{equation}
while the temporal pairs couple to a gauge field $U(t)$ and internal interactions via
\begin{equation}
  E_i \;\longrightarrow\; E_i - U(t_i) - \Omega_i^{\mathrm{int}}(\bm t).
\end{equation}
The resulting many-body Carroll generator is
\begin{equation}
  \mathcal H^{(N)}_{\mathrm{in}}[\Phi,U]
  =
  \sum_{i=1}^N \frac{1}{2mc^2}
  \Big(\widehat E_i - \Omega_i^{\mathrm{int}}(\bm t) - U(t_i)\Big)^{2}
  \;+\;
  \sum_{i=1}^N \Phi(t_i),
  \qquad
  \widehat E_i = -i\hbar\,\partial_{t_i},
  \label{eq:Carroll-H-in-DFT}
\end{equation}
where $\Omega_i^{\mathrm{int}}$ encodes fixed internal interactions. Under the substitutions
\begin{equation}
  q \leftrightarrow t, \quad
  \hat{p} \leftrightarrow \widehat E/c, \quad
  v(q) \leftrightarrow \Phi(t), \quad
  A(q) \leftrightarrow U(t),
\end{equation}
where $q$ is a 1D spatial coordinate, this problem is formally isomorphic to standard \emph{one-dimensional} current-density functional theory (CDFT) \cite{Lieb1983,VignaleRasolt1987,VignaleRasolt1988,EngelDreizler2011}. As in 1D CDFT, $U(t)$ is defined only up to gauge transformations $U\to U+\partial_t\chi$ accompanied by $\Psi\to e^{\frac{i}{\hbar}\sum_i \chi(t_i)}\Psi$, while $\Phi(t)$ remains gauge invariant.

We import the existence and uniqueness theorems of CDFT without rederivation, assuming the Hamiltonian is bounded from below, the ground state is nondegenerate up to gauge, and $v$-representability holds. Thus, for fixed $\Omega_i^{\mathrm{int}}$, there exists a universal functional $F[n,j_t]$ of the temporal density $n(t)$ and physical current $j_t(t)$. Universality implies independence from $(\Phi, U)$, though $F$ depends parametrically on the internal interaction.

The interacting densities are reproduced by a fictitious noninteracting system via one-time orbitals $\{\varphi_k(t)\}$ solving the Carroll--Kohn--Sham equations
\begin{equation}
  \frac{1}{2mc^{2}}
  \Big(-i\hbar\partial_t - U_s[n,j_t]\Big)^{2}\varphi_k
  \;+\; \Phi_s[n,j_t]\,\varphi_k
  = \epsilon_k\,\varphi_k,
  \label{eq:Carroll-KS-eq}
\end{equation}
with
\begin{equation}
  n(t) = \sum_k f_k\,|\varphi_k|^{2},
  \quad
  j_t(t)
  = \frac{1}{mc^{2}}
    \sum_k f_k\,
    \mathrm{Re}\!\left\{
      \varphi_k^{*}\big(-i\hbar\partial_t - U_s\big)\varphi_k
    \right\}.
\end{equation}
The effective fields $\Phi_s$ and $U_s$ include external, Hartree, and exchange--correlation contributions determined by the functional derivatives of $F[n,j_t]$. This framework enables the analysis of collective temporal behavior and emergent many-body phenomena using the established language of spatial DFT.

\section{Discussion}

In this work we developed a many-body Carroll--Schr\"odinger framework in which the spatial variable $x$ acts as an evolution parameter and the temporal variables $\bm t=(t_1,\ldots,t_N)$ play the role of configuration coordinates. Starting from a relativistic multi-time Klein--Gordon model and performing a Carrollian contraction, we obtained a consistent equal-$x$ generator that is a sum of temporal energy operators acting on $L^2(\mathbb R^N_{\bm t})$. Temporal interactions were introduced via minimal coupling, and an $x$-dependent gauge transformation mapped these couplings to explicit many-body potentials in the time variables, providing a systematic route to construct Carrollian partners of familiar Schr\"odinger models.

The analysis of spatial potentials revealed a characteristic Carrollian kinematics. When the starting point is a static spatial potential $U_{\!tot}(\mathbf x)$, the transformed CS evolution depends on $U_{\!tot}$ only through the net collective force $\sum_j\partial_{x_j}U_{\!tot}(\mathbf x)$. For coupled harmonic oscillators this produces a calculable linear-in-$t$ drive of the collective coordinate and explicit Gaussian $N$-body solutions. For any spatially translation-invariant interaction (including regularized one-dimensional Coulomb potentials), the internal forces cancel in the collective direction, so that the collective Carrollian dynamics becomes free and the potential reappears only through phases and boundary data. From the spatial viewpoint this yields a class of spatially homogeneous, boundary-data–determined evolutions without time-translation symmetry, reminiscent of homogeneous fluids whose dynamics is entirely fixed by their initial conditions and a global time-dependent drive. In this sense the construction illustrates, in a simple many-body setting, how Carrollian ultralocality enforces consistency of spatially driven systems with their initial data.

On the operator side, we generalized the single-particle coordinate duality between the time-independent Schr\"odinger equation and a space-independent CS equation to separable many-body systems using Schwarzian relations for the inverse coordinate maps. For each particle a reparametrization $x_i=\delta_i(t_i)$ relates static Schr\"odinger potentials $V_{\mathrm{sch}}^{(i)}(x_i)$ to one-body Carroll potentials $V_{\mathrm{car}}^{(i)}(t_i)$ acting on the temporal coordinates. We also explored this Schwarzian map in an interacting setting, namely two coupled Schr\"odinger oscillators, which suggests that certain classes of separable interacting models can be transplanted into the Carrollian framework. Exchange symmetry was formulated directly in the time domain, leading to bosonic and fermionic sectors characterized by temporal bunching or antibunching through the second-order coherence function $g^{(2)}(t,t')$ on equal-$x$ slices.

In a second-quantized description on equal-$x$ slices, a two-time interaction kernel leads, in a short-memory limit analogous to the contact-interaction regime of the Lieb--Liniger gas, to an structure similar to a temporal derivative cubic--quintic nonlinear Schr\"odinger equation. In this limit the structure of the minimal coupling fixes the quintic coefficient to $\beta= -3/16$, placing the interaction-driven temporal Carroll gas in a specific universality class of derivative nonlinear Schr\"odinger systems. This provides a concrete example where Carrollian kinematics constrains the effective nonlinear dynamics of a many-body system in the time domain.

Finally, we identified a natural mapping between the static CS many-body problem and one-dimensional current-density functional theory (CDFT). By minimally coupling the canonical momentum $P$ to a scalar field $\Phi(t)$ and the temporal energy $E$ to a gauge field $U(t)$, the ground-state problem on equal-$x$ slices can be written in a form that is formally isomorphic to one-dimensional CDFT. Rather than proving new existence theorems, we use this isomorphism to motivate a universal functional $F[n,j_t]$ of the temporal density $n(t)$ and temporal current $j_t(t)$ and to outline a Carrollian Kohn--Sham scheme in terms of one-time orbitals that reproduce $(n,j_t)$ on equal-$x$ slices.

The present analysis, while opening the door to multiparticle Carrollian quantum mechanics, is necessarily restricted. We have worked in $1{+}1$ dimensions, considered spinless particles and specific classes of temporal and spatial potentials, and focused our explicit solutions on Gaussian states and exactly solvable models (such as coupled oscillators and the trivialized Coulomb gas). The coordinate duality was implemented for separable Schr\"odinger potentials and explored in an interacting example of two coupled harmonic oscillators; extending it to general $N$-body interacting static potentials would require genuine multivariable generalizations of the Schwarzian map and remains an open problem. Likewise, our discussion of Carrollian density-functional theory has been kept at a formal level: concrete temporal exchange--correlation approximations were not attempted here and would have to be adapted from the CDFT literature or reformulated to take into account the temporal behaviour of Carrollian particles.

Despite these limitations, the results illustrate that many-body Carroll--Schr\"odinger dynamics, coordinate dualities, temporal coherence and functional descriptions can be organized within a common framework. Beyond the intrinsic interest of Carrollian quantum systems, there are several potential areas of application. Carroll symmetry has already appeared in cosmology and dark-energy model building, where Carrollian fluids naturally realize an equation of state $\mathcal E + P = 0$ and provide a useful language for inflation and de Sitter-like phases~\cite{deBoer2022,deBoerStories}. In parallel, Carrollian geometry and hydrodynamics have been linked to effective descriptions of strongly driven or constrained systems, including Carrollian fluids and Bjorken flow, and other condensed-matter models with restricted mobility~\cite{BagchiBjorken,CarrollHydro,MarsotHall}. The spatially driven multiparticle Carroll systems studied here, with their homogeneous collective response and strong dependence on boundary data, may provide simple toy models for such homogeneous yet nonstationary Carrollian media.

A natural next step is to generalize the present construction to higher spatial dimensions and to more elaborate Carrollian field theories. The Carroll--Schr\"odinger equation itself admits higher-dimensional and gauge-coupled versions, both as the ultra-relativistic limit of tachyonic models and as field theories with nontrivial Carroll--Schr\"odinger symmetry algebras~\cite{Najafizadeh2025-2}. At the same time, two-time formulations of Carroll particles suggest that additional temporal and spatial dimensions can be incorporated within a unified constrained Hamiltonian framework~\cite{Kamenshchik2024}. Extending the many-body CS formalism along these lines, and combining it with temporal density functionals and DNLS-type effective theories, may open a bridge between Carrollian quantum mechanics, condensed-matter models with emergent Carroll symmetry, and cosmological applications such as dark energy dynamics.

\end{document}